\documentclass[twocolumn]{emulateapj}

\makeatletter
\newcommand*{\rom}[1]{\expandafter\@slowromancap\romannumeral #1@}
\makeatother

\def\fl[#1 #2]{{#1}$\pm${#2}}
\def\orb[#1 #2]{{$#1^{#2}$}}
\def\ls[#1 #2]{{$^{#1}${#2}}}
\def\tm[#1 #2 #3]{{$^{#1}${#2}$_{#3}$}}
\def\ion[#1 #2]{#1\,{\sc #2}}

\begin{document}

\submitted{Submitted to the Astrophysical Journal}
\journalinfo{Submitted to the Astrophysical Journal}
\title{EUV emission and scattered light diagnostics of equatorial coronal holes as seen by \textit{Hinode}/EIS}

\author{Carolyn Wendeln$^{1}$, Enrico Landi$^{1}$}
\altaffiltext{1}{Department of Climate and Space Sciences and Engineering, University of Michigan, Ann Arbor MI 48109, USA}
\begin{abstract}

Spectroscopic diagnostics of solar coronal plasmas critically depends on the uncertainty in the measured line intensities. One of the main sources of uncertainty is instrumental scattered light, which is potentially most important in low-brightness areas. In the solar corona, such areas include polar and equatorial coronal holes, which are the source regions of the solar wind; instrument-scattered light must thus pose a significant obstacle to studies of the source regions of the solar wind. In this paper we investigate the importance of instrument-scattered light on observations of equatorial coronal holes made by the Hinode/EIS spectrometer in two different phases of the solar cycle. We find that the instrument-scattered light is significant at all temperatures, and in both regions it amounts to approximately 10\% of the average intensity of the neighboring quiet Sun regions. Such contribution dominates the measured intensity for spectral lines formed at temperatures larger than Log T = 6.15 K, and has deep implications for spectroscopic diagnostics of equatorial coronal hole plasmas and studies of the source regions of a large portion of the solar wind which reaches Earth. Our results suggest that the high temperature tail in the coronal hole plasma distribution with temperature, however small, is an artifact due to the presence of scattered light.

\end{abstract}


\section{Introduction}\label{intro}

Coronal holes are known as funnel-like regions of rapidly expanding open magnetic field lines. They are commonly accepted as the large-scale source region of the fast solar wind, and are characterized by both a lower electron temperature and density than the adjacent quiet Sun closed field regions \citep{1973SoPh...29..505K,1977RvGSP..15..257Z}. Compared to coronal holes, quiet Sun regions are occupied by only a small fraction of coronal funnels \citep{2001A&A...374.1108P}

In order to better understand both the source region and acceleration mechanism of the fast solar wind it is imperative to measure both the thermal structure and physical properties of coronal hole plasmas.
The extreme-ultraviolet (EUV) wavelength range is particularly useful for this task because it is rich in strong spectral lines that are suitable for plasma diagnostics \citep{2007PASJ...59S.857Y}. High-resolution spectroscopic analysis of EUV emission is thus a vital tool for investigating the physical properties of plasma.

Plasma diagnostics techniques can be used to measure the electron and ion temperatures, electron density, and the thermal structure of the plasma from line intensities of spectra. In addition, some diagnostics techniques use the measured full width at half-maximum (FWHM) to measure ion temperatures and estimate non-thermal motions. The knowledge of such plasma properties is important for modeling small-scale transient phenomena, as well as the understanding of their role in coronal heating and solar wind generation \citep[e.g.,][]{2010ApJ...709..499S,Dudík2017}.

The thermal structure of coronal spectra is most often inferred using a Differential Emission Measure (DEM) curve. DEM determinations aid in providing an accurate estimates of the thermodynamic state of coronal hole plasmas at varying heights. However, all these measurements are negatively affected by a number of uncertainties. As discussed by \citet{2011ApJ...736..101H}, uncertainties can arise from the diagnostic method, measured intensities, as well as from the atomic data and plasma parameters. One such uncertainty is instrumental scattered light.

Scattered light is the part of radiation which strikes the CCD detector following paths that were not originally intended. Scattered light originates from several sources, the most important being diffraction effects due to the mesh supporting the filters along the optical path, to scattering by the reflecting surface (the multiplayer coating), by electron diffusion in the CCD, and to other smaller effects. A summary of such effects can be found in \citet{aiapsf} and \citet{2013ApJ...765..144P}. These effects determine the instrumental Point Spread Function (PSF). However, the brightness of the target determines the spatial distribution of the intensity in the field of view (FOV), so that the same PSF provide different amounts of scattered light depending on the particular target being observed.

The importance of scattered light has triggered several studies that attempt to determine the instrumental PSF for the EUV narrow-band imagers currently operating. \citet{aiapsf} and \citet{2013ApJ...765..144P} utilized instrumental parameters measured before launch and lunar transit on the Sun to determine two independent estimates of the PSF on each of the 7 coronal channels of the SDO/AIA instrument. \citet{2012ApJ...749L...8S}, instead, applied a blind deconvolution technique to a lunar transit on the Sun to determine the PSF of the EUVI instrument on board STEREO-B. They further estimated that scattered light provides 40\%-70\% of the measured intensity in equatorial coronal holes.

In order to estimate the amount of stray light contamination for the EUV Imaging Spectrograph \citep[EIS,][]{2006ApOpt..45.8674K,2007SoPh..243...19C} on board \textit{Hinode} \citep{2007SoPh..243....3K}, \citet[-- hereafter UU10]{stray} used an observation of a partial solar eclipse of the Sun as seen by \textit{Hinode}. UU10 tried to separate the stray light component from the direct component by aiming the detector at the (eclipsed) solar disk when no direct light should be striking the detector. They subtracted the dark currents from their observation so that all that remained was the instrumental scattered light. They found the scattered light for EIS emission to be a minimum of 2\% of the average on-disk emission at a given wavelength.

It has become standard a practice to subtract 2\% of the average on-disk emission from the intensities observed in the off-disk portion of the Sun \citep[e.g.,][]{2011ApJ...736..101H,2012ApJ...751..110B,2012ApJ...757...73K}. \citet{2011ApJ...736..101H} checked this method and found it to be reasonable by analyzing on- and off-disk data from an He II, Si X blend at 256.3 {\AA} and Si X at 261.0 {\AA}. However, as noted by \citet{2012ApJ...757...73K}, the stray light amount computed by UU10 may be an underestimate.

The observation taken by UU10 was on-disk during a partial eclipse; however, since less than half of the Sun's disk was illuminated it resembled an off-limb observation. Since the amount of scattered light also depends on the intensity distribution in the instrument FOV, the stray light correction of UU10 might be an underestimate for EIS on-disk observations when the entire FOV is emitting. Coronal holes are more susceptible to this underestimate of scattered light since they are much cooler and dimmer than the surrounding quiet Sun. To date there has not been a study carried out which addresses the potential impact of scattered light on plasma diagnostics from disk observations of coronal holes.

The goal of this paper is to provide an estimate of scattered light for on-disk observations of equatorial coronal holes, and to discuss its impact on plasma diagnostics and on the plasma properties in equatorial coronal holes. In \S 2 we describe the observations we have chosen, in \S 3 we determine the appropriate level of scattered light from our observations, in \S 4 we estimate the expected emission of the corona, and in \S 5 we discuss our findings.

\section{Observations and Data Analysis}

EIS takes high resolution spectra in two wavelength bands: 170-212 {\AA} (short wavelength band or SW) and 246-292 {\AA} (long wavelength band or LW). These two wavelength bands include a large number of emission lines that offer excellent diagnostic opportunities for coronal plasmas. Uncertainties in the measured intensities (i.e., scattered light) can easily contaminate these plasma diagnostics for both coronal holes and the adjacent quiet Sun.

\begin{table}[]
\caption{This table shows all of the lines which are used in the present analysis. Formation temperatures, as well as the standard wavelengths, are also listed. The ratios of average coronal hole to average quiet Sun intensities are listed for each line of each data set (see text for details).}
\centering 
\begin{tabular}{c c c c c c} 
\hline\hline 
\\
Ion & Log T (K) & \multicolumn{2}{c}{2007 March 31} & \multicolumn{2}{c}{2013 October 12} \\
& & $\lambda$ ({\AA}) & Ratio & $\lambda$ ({\AA}) & Ratio \\ [0.5 ex] 
\\
\hline 
\\\
\ion[He ii] & 4.90 & 256.32 & 0.58 & 256.32 & 0.55 \\
\\
\ion[Fe viii] & 5.64 & 185.21& 0.76 & 186.59 & 0.61 \\
\\
\ion[Si vii] & 5.78 & 275.37 & 0.82 & 275.37 & 0.94\\
\\
\ion[Fe ix] & 5.91 & 197.86 & 0.49 & 197.86 & 0.55\\
\\
\ion[Fe x] & 6.04 & 177.23 & 0.37 & 184.54 & 0.33 \\
& & 257.27 & 0.41 & & \\
\\
\ion[Fe xi] & 6.13 & 180.40 & 0.20 & 180.40 & 0.15 \\
& & 182.17 & 0.21 & & \\
& & 256.93 & 0.22 & & \\
\\
\ion[Si x] & 6.15 & 258.38 & 0.15 & 258.38 & 0.17 \\
\\
\ion[S x] & 6.17 & 264.24 & 0.14 & 264.24 & 0.15 \\
\\
\ion[Fe xii] & 6.19 & 195.12 & 0.12 & 195.12 & 0.12 \\
& & 193.51 & 0.12 & & \\
\\
\ion[Fe xiii] & 6.25 & 202.05 & 0.11 & 202.05 & 0.10 \\
\\
\ion[Fe xiv] & 6.29 & 264.79 & 0.15 & 264.79& 0.12 \\
\\
\ion[Fe xv] & 6.35 & 284.17 & 0.15 & 284.17 & 0.11 \\
\\
\hline \\
\end{tabular}
\label{table:nonlin} 
\end{table}

In order to determine the amount of scattered light which affects coronal holes as seen by EIS we need observations of equatorial coronal holes which have a high signal to noise and a FOV large enough to include both coronal hole and the surrounding quiet Sun. Equatorial coronal holes are used for this study because they receive the maximum amount of illumination from the surrounding Sun when compared to polar coronal holes. In addition, there are less projection effects caused by overlaying closed field, denser structures. There are only a few data sets which meet this criteria, nevertheless we have selected two which will provide excellent analysis for our study. We have inspected many other data sets in the EIS archive, but they usually either did not include the whole coronal hole, or - most frequently - their signal to noise was too low to allow an accurate determination, and were discarded.

Our primary data set is an EIS raster obtained on 2007 March 31 on an equatorial
coronal hole close to disk center, and centered at around (500",180"). This raster
was made of 14 individual rasters, each 14" $\times$ 512" wide, each obtained moving
the 2" slit at 7 adjacent positions along the E-W direction. The pointing of these
14 rasters was calculated to cover continuously a 250"$\times$512 field of view, 
which included almost entirely the equatorial coronal hole and the surrounding quiet
Sun areas. The exposure time was 300 s for each slit position; the entire length of
the EIS slit, and the whole wavelength ranges covered by the two EIS detectors were
downloaded to the ground, resulting in a complete EIS spectrum at each spatial pixel 
of the field of view. Unfortunately, two individual rasters were lost, leaving a
gap close to the center of the field of view. The wide slit and the long exposure
time ensured an excellent signal-to-noise ratio.


\begin{figure*}
\begin{center}
\includegraphics[width=3.5in,angle=180]{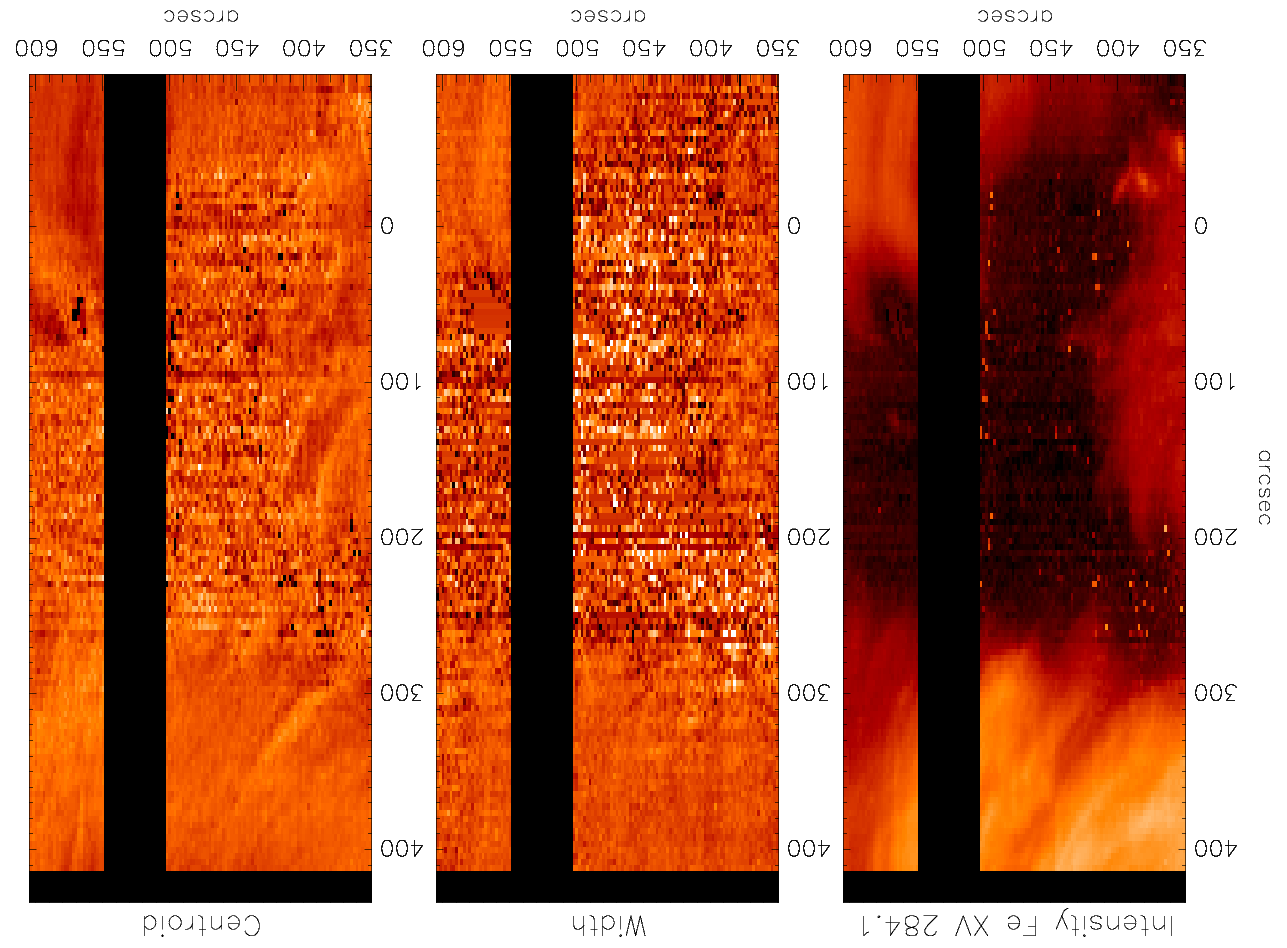}
\includegraphics[width=3.5in,angle=180]{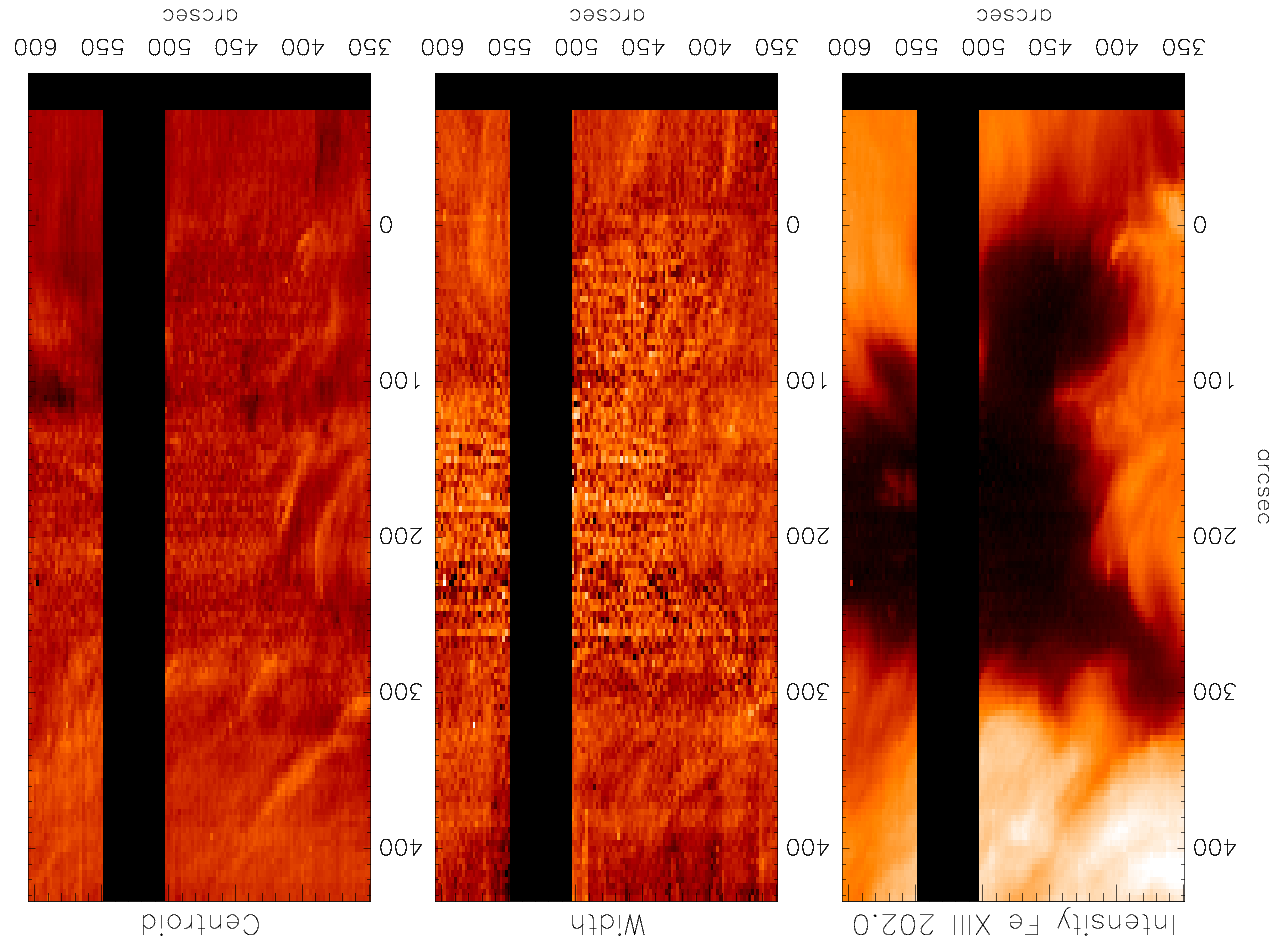}
\includegraphics[width=3.5in,angle=180]{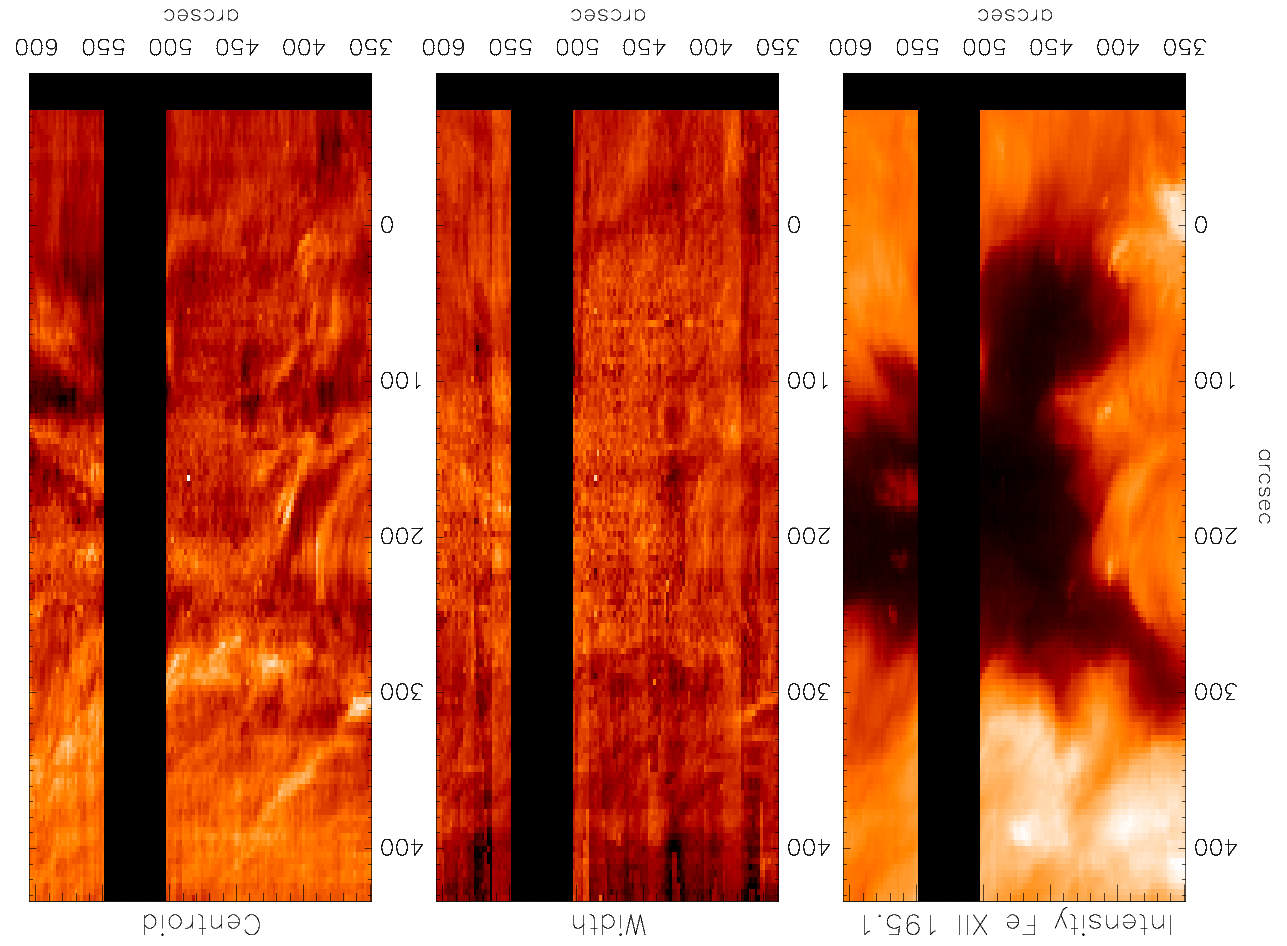}
\includegraphics[width=3.5in,angle=180]{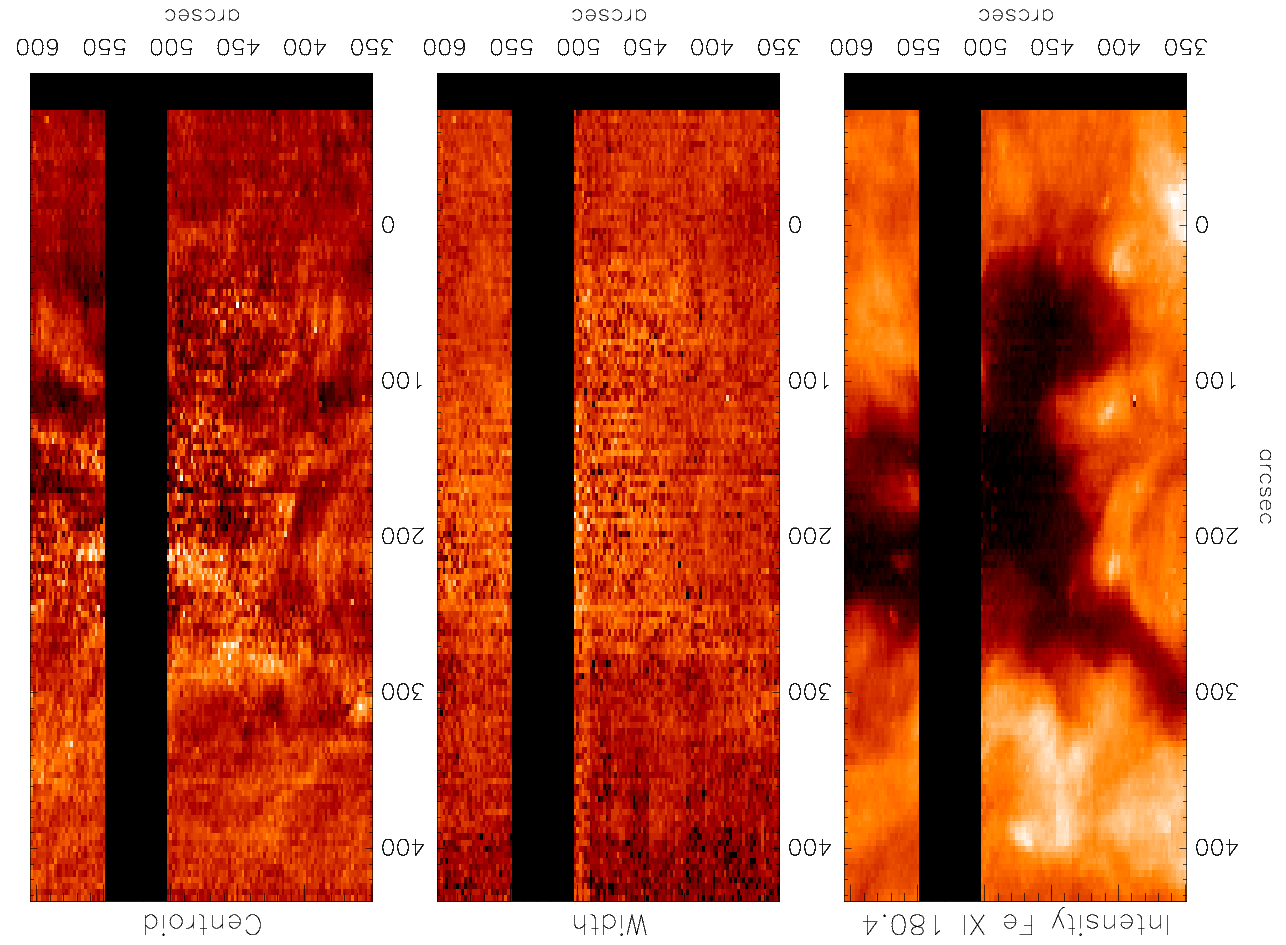}
\includegraphics[width=3.5in,angle=180]{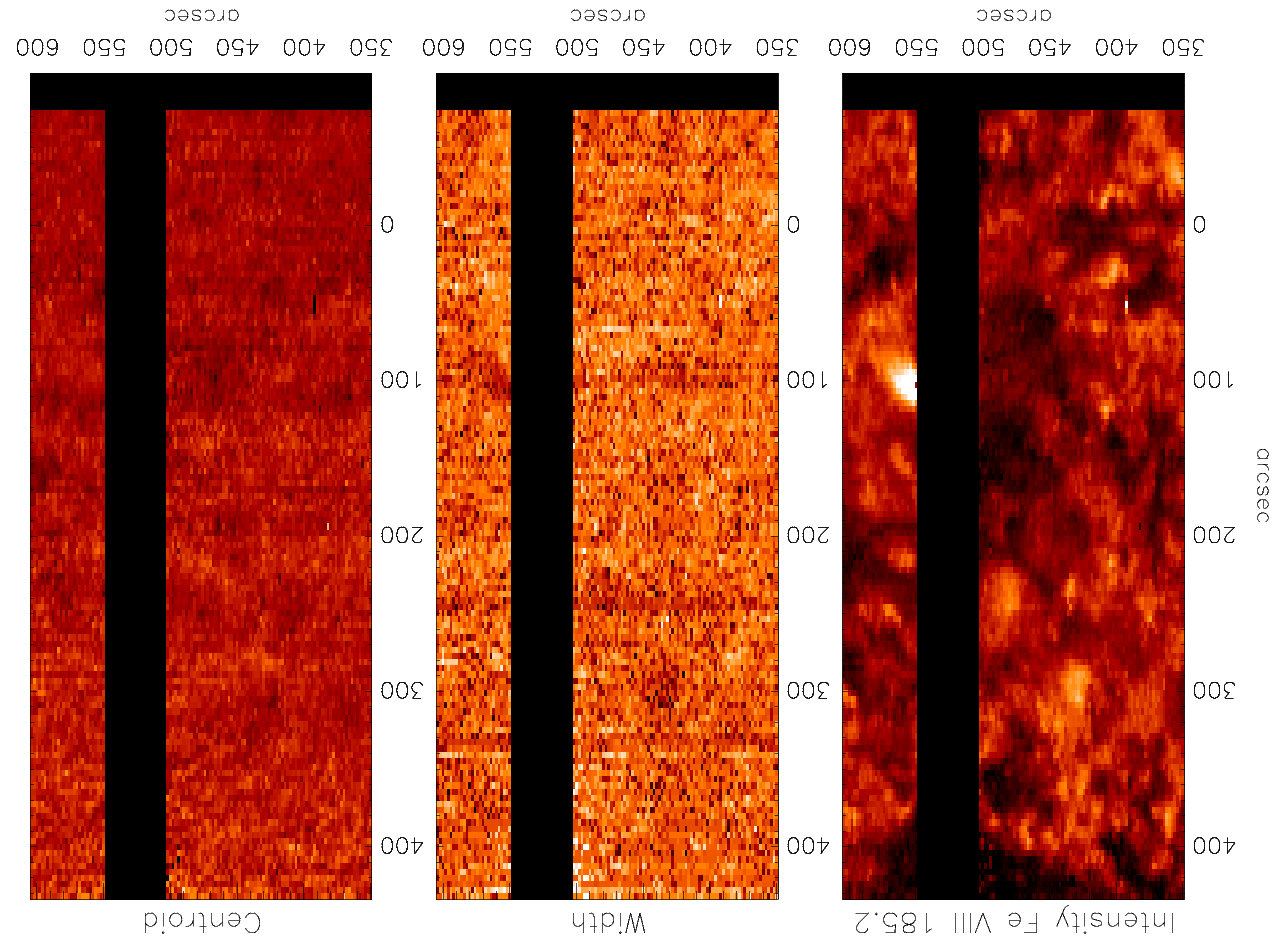}
\includegraphics[width=3.5in,angle=180]{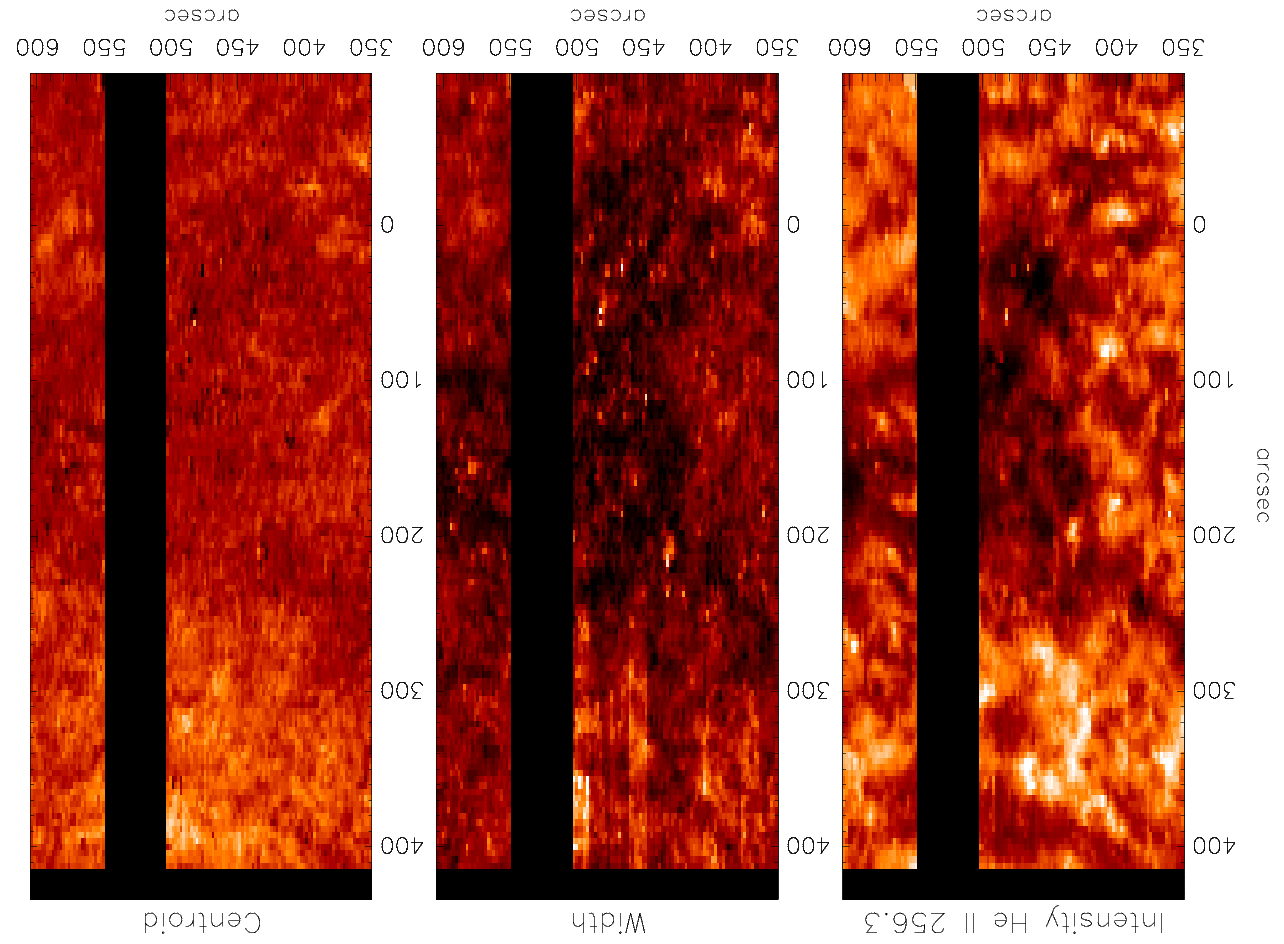}
\caption{Intensity, width, and centroid maps for \textbf{Top Left:} \ion[Fe xv] at 284.17 {\AA} \textbf{Top Right} \ion[Fe xiii] at 202.05 {\AA} \textbf{Middle Left} \ion[Fe xii] at 195.12 {\AA} and \textbf{Middle Right} \ion[Fe xi] at 180.40 {\AA} \textbf{Bottom Left} \ion[Fe viii] at 185.21 {\AA} and \textbf{Bottom Right} \ion[He ii] at 256.32 {\AA}}
\end{center}
\end{figure*}

The second data set was taken on 2013 October 12 of another equatorial
coronal hole near disk center. This data set consisted in a single raster
obtained by rastering the 1" slit along the E-W direction with a 2" step.
The final field of view was 240"$\times$512" because the entire slit was
transmitted to the ground. However, only a few selected spectral windows 
were downloaded, which included bright isolated lines from \ion[Fe viii] to \ion[Fe xv],
\ion[He ii], \ion[S x] and \ion[Si vii]. At each slit position, the exposure 
time was 100 s; the shorter exposure time and the narrower slit caused the
signal-to-noise ratio of this data set to be lower than in the primary data
set by a factor $\approx$2.5. The relatively small size of the coronal hole 
allowed this raster to include most of the coronal hole and
some adjacent quiet Sun. Also, a very small bright point was located
at the edge of the coronal hole, which allowed us to investigate
whether its presence had any effect on the coronal hole intensity in
its proximity. Since this data set has a lower signal-to-noise, we will 
use it to check whether or not the results from our primary data set are 
consistent (even with varying location on the Sun, year, and type of 
observation).

\begin{figure}
\begin{center}
\includegraphics[width=8.5cm,angle=180]{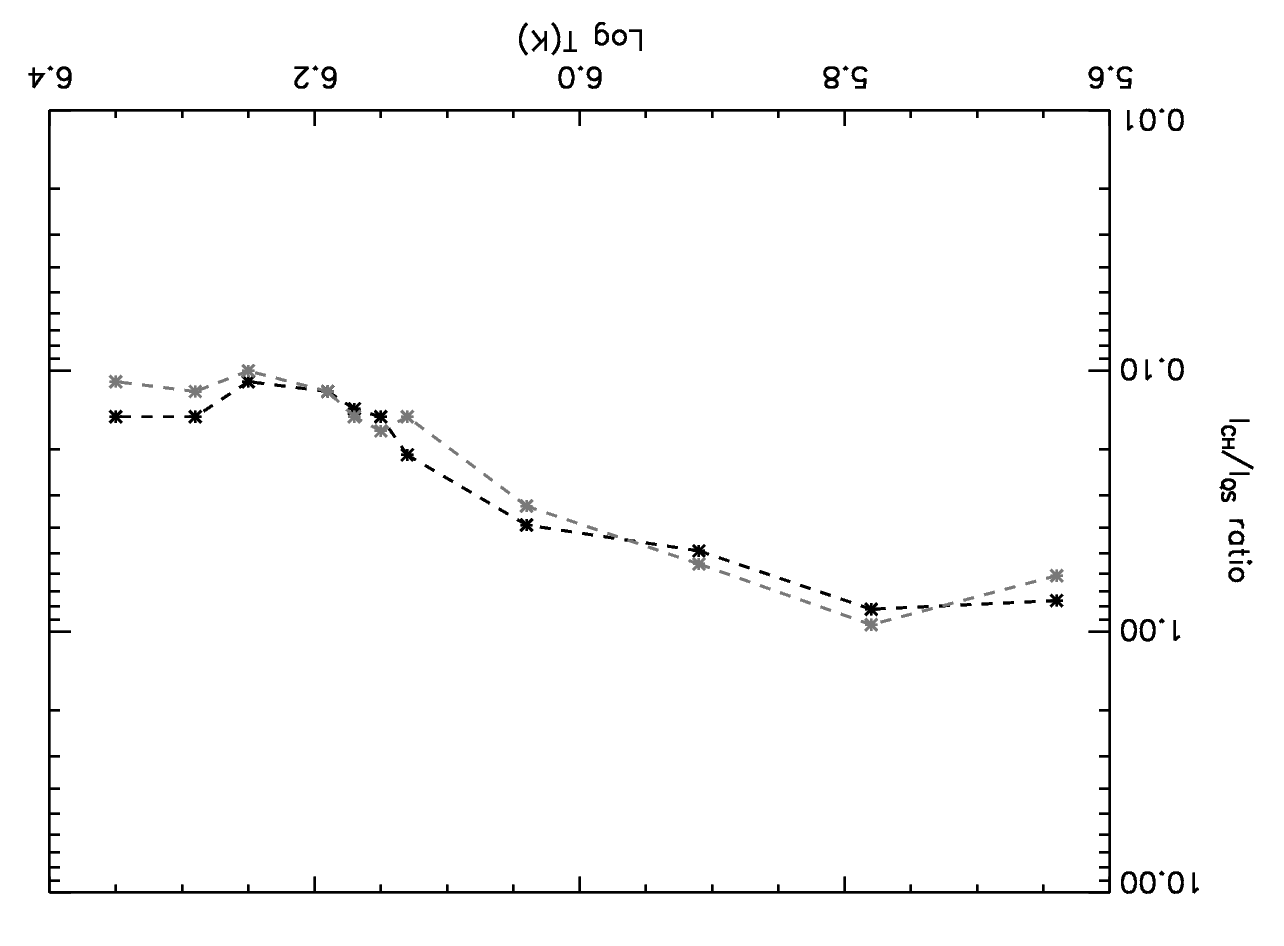}
\caption{
Observed intensity ratio for the ions listed in Table 1.
The black line shows the ratio for our primary data set and the grey line shows the ratio for our secondary data set.}
\end{center}
\end{figure}

The raw EIS spectra were first cleaned and calibrated using standard procedures available in SolarSoft (SSW). The routine \tt ${EIS\_PREP}$ \rm removes the CCD dark current, cosmic ray strikes on the CCD, and applies a flat field correction. In addition, it flags and removes hot, warm, and dusty pixels and performs an absolute calibration to convert the data from photon events to physical units of erg cm$^{-2}$ s$^{-1}$ sr$^{-1}$ $\AA^{-1}$. Lines from 12 ions (sometimes multiple lines per ion) have been selected for this present analysis and are listed in Table 1. These lines were chosen because they are all strong and isolated lines which are in spectral regions which are relatively free of other lines so the background is very easy to determine. Self-blended lines are not used in this study for their increased uncertainties reach unacceptable limits for automatic Gaussian fits. \ion[He ii] is a blend of several different coronal lines, however it dominates over other spectral lines in disk spectra \citep{2009A&A...495..587Y,2012A&A...538A..88G,2015ApJ...803...66G} and therefore is acceptable to use in our study.

To improve the accuracy of the centroid and width measurements for our primary data set, the data were binned by a factor of 4 in the vertical direction using the routine \tt ${EIS\_BIN\_WINDATA}$ \rm. In addition, our secondary data set was binned by a factor of 2 in both the horizontal and vertical directions. Gaussians were then fitted to this binned data to extract the emission line intensities using the routine \tt ${EIS\_AUTO\_FIT}$\rm. This routine automatically fits Gaussians to two-dimensional spatial arrays of \textit{Hinode}/EIS spectra and returns arrays for intensity, centroid, and width, in addition to their associated 1-$\sigma$ error arrays.

\section{Observed Intensities in Equatorial Coronal Holes}

\begin{figure}[t]
\begin{center}
\includegraphics[width=4in,angle=0]{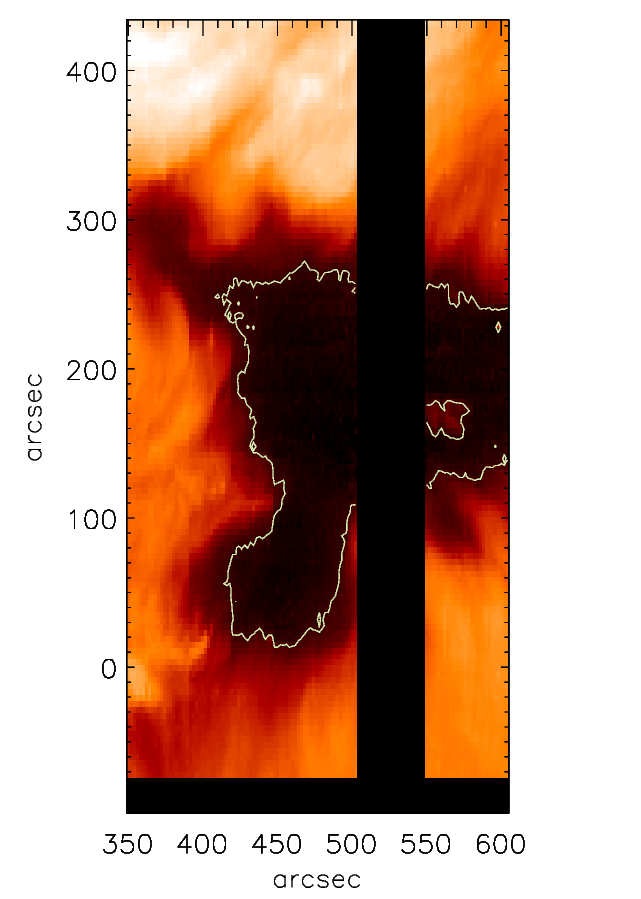}
\caption{The mask used for separating the coronal hole from the quiet Sun using \ion[Fe xiii] at 202.05 {\AA} for our primary data set on 2007 March 31. Every intensity value above 22.4 erg/cm$^{2}$/s/sr (seen as the yellow contour line) is defined to be the quiet Sun region and everything below is defined to be the coronal hole.}
\end{center}
\end{figure}

Figure 1 shows maps of intensity, centroid, and width for individual spectral lines from our main dataset. We report two chromosphere/ transition region lines ( \ion[Fe viii] at 185.21 {\AA} and \ion[He ii] at 256.32 {\AA}), two lines typically formed in the non-active corona ( \ion[Fe xi] at 180.40 {\AA} and \ion[Fe xii] at 195.12 {\AA}), and two lines that can be found both in quiet Sun and non-flaring active regions (\ion[Fe xv] at 284.17 {\AA} and \ion[Fe xiii] at 202.05 {\AA}).

The spectral lines \ion[Fe viii] at 185.21 {\AA} and \ion[He ii] at 256.32 {\AA} form in the upper transition region and in the upper chromosphere, respectively. The intensity maps of the coronal hole for these two lines appear smaller and more structured compared to the hotter ions. The hottest line, \ion[Fe xv] at 284.17 {\AA}, loses the majority of its structure within the coronal hole and appear to be flattened out by noise. This pattern is also reflected in the centroid and width maps - despite the fact that these maps are inherently more noisy than the intensity maps. For all of our hotter spectral lines listed in Table 1 (i.e., above \ion[Fe x]) the structure of the coronal hole appears to be flat and featureless, while there is still visible structure seen within the coronal hole for \ion[He ii] and \ion[Fe viii]. A similar behavior is also found in the \ion[Fe xiii] 202.05 {\AA} line, which is also formed at high temperatures. As far as the quiet Sun lines (\ion[Fe xi] and \ion[Fe xii]) are concerned, their intensities, widths, and centroids still show some structure, which is more evident in \ion[Fe xi] and is almost lost in \ion[Fe xii].

\begin{figure}
\begin{center}
\includegraphics[width=3.6in,angle=0]{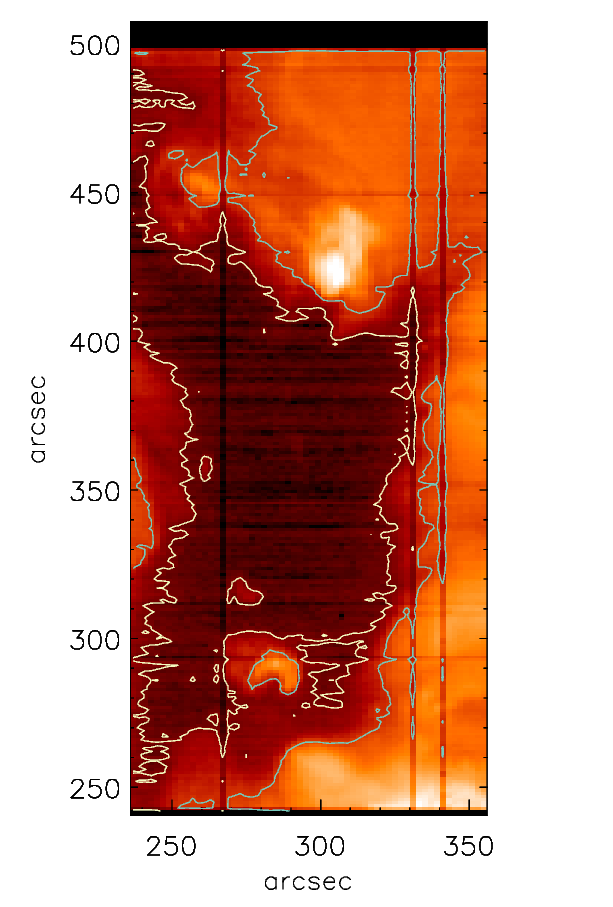}
\caption{The mask used for separating the coronal hole from the quiet Sun using \ion[Fe xiii] at 202.05 {\AA} for our secondary data set on 2013 October 12. Every intensity value above 700 erg/cm$^{2}$/s/sr (seen as the blue contour line) is defined to be the quiet Sun region and every intensity value below 200 erg/cm$^{2}$/s/sr (seen as the yellow contour line) is defined to be the coronal hole. }
\end{center}
\end{figure}

The apparent uniformity and lack of structure seen in the intensity, centroid, and width measurements of the coronal hole (as opposed to the quiet Sun) begs the question of whether this feature is due to intrinsic lack of structures in the emitting plasma at high temperature or if it is due to the presence of instrument scattered light. The latter should display a far smoother spatial distribution throughout the image. In addition, the intensity, centroid, and width maps should appear flat since it would be attributed to an average of the surrounding emission made by the instrument. The flat structure of the coronal hole seen in the maps for both \ion[Fe xv] at 284.17 {\AA} and \ion[Fe xiii] at 202.05 {\AA} is consistent with the expected characteristics of scattered light.

A potential source of scattered light for our data sets may come from the quiet Sun region surrounding the coronal holes. Since the centroid and width results in Figure 1 are too noisy we will focus on our intensity measurements for the remaining portion of our study. In order to see how the intensity measurements vary between the coronal hole and quiet Sun we will separate the FOV between these two regions and average out their intensities in order to study the behavior of the intensity ratio between the coronal hole and quiet Sun.

Figure 2 shows the ratio of the median coronal hole intensity divided by the median of the quiet Sun intensity for each line listed in Table 1. We have omitted \ion[He ii] from this Figure (and from subsequent figures) because this line is optically thick. See Section 4.3 for results. For our primary data set a single mask was applied to the FOV of each ion using \ion[Fe xiii] at 202.05 {\AA} (see Figure 3) as a template to separate the coronal hole from the quiet Sun. \ion[Fe xiii] at 202.05 {\AA} is an excellent line for showing a distinct boundary between the coronal hole and the quiet Sun. A contour map was applied to the intensity map of \ion[Fe xiii] at 202.05 {\AA} until a level was found that adequately separated the coronal hole from the quiet Sun. Every intensity value above 22.4 erg/cm$^{2}$/s/sr is associated to the quiet Sun region and everything below was associated to be the coronal hole. Figure 4 shows the masks applied to the FOV of our secondary data set. Every intensity value above 700 erg/cm$^{2}$/s/sr (seen as the blue contour line) is defined to be the quiet Sun region and every intensity value below 200 erg/cm$^{2}$/s/sr (seen as the yellow contour line) is defined to be the coronal hole. The boundary values for coronal hole and quiet Sun were chosen in each data set empirically, as the values that allowed us to best separate the two regions. The values are very different for the two data set, this can be due to the fact that these two regions have been observed several years apart: the 2007 one at the beginning of the deep minimum between cycles 23 and 24, the 2013 one close to solar maximum. The different phase of the solar cycle might have affected the brightness levels of both types of region.

\begin{figure*}
\begin{center}
\includegraphics[width=3.45in,angle=0]{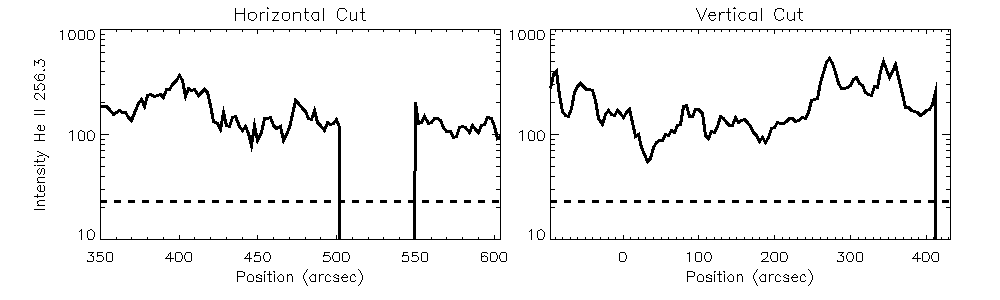}
\includegraphics[width=3.45in,angle=0]{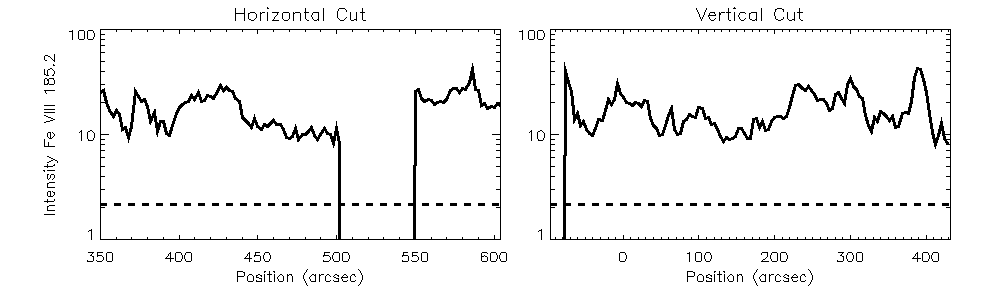}
\includegraphics[width=3.45in,angle=0]{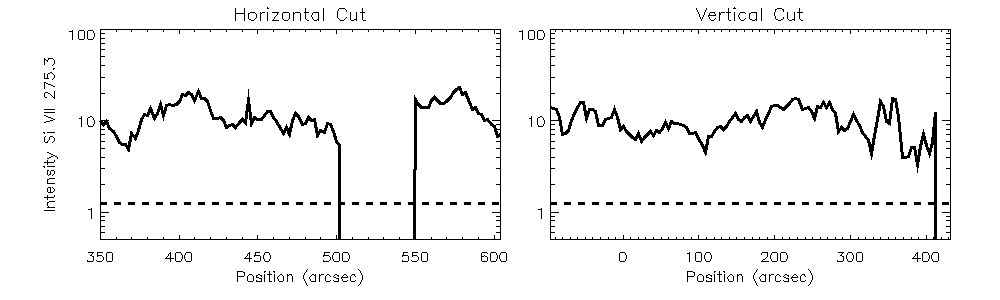}
\includegraphics[width=3.45in,angle=0]{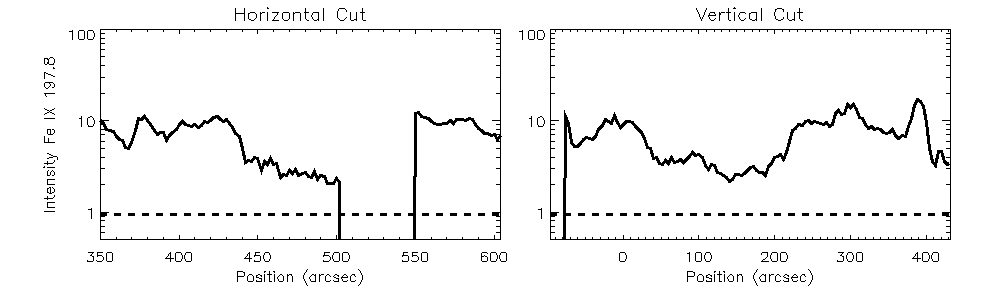}
\includegraphics[width=3.45in,angle=0]{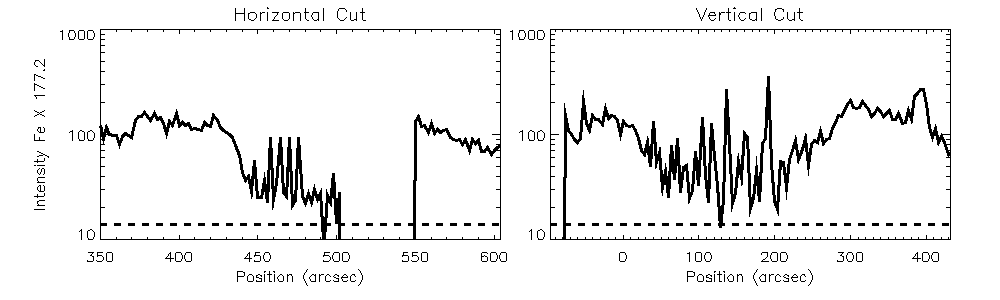}
\includegraphics[width=3.45in,angle=0]{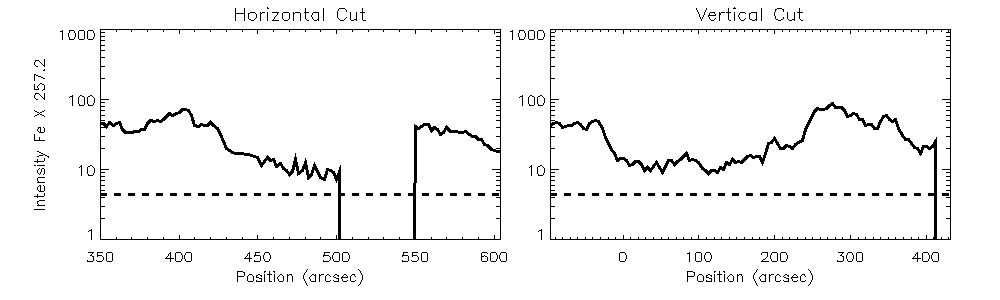}
\includegraphics[width=3.45in,angle=0]{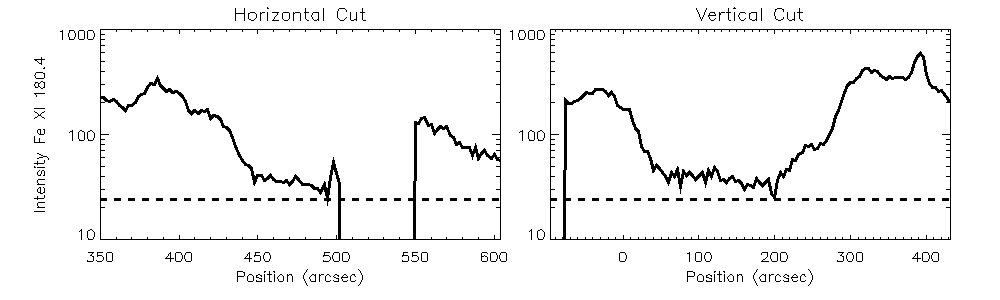}
\includegraphics[width=3.45in,angle=0]{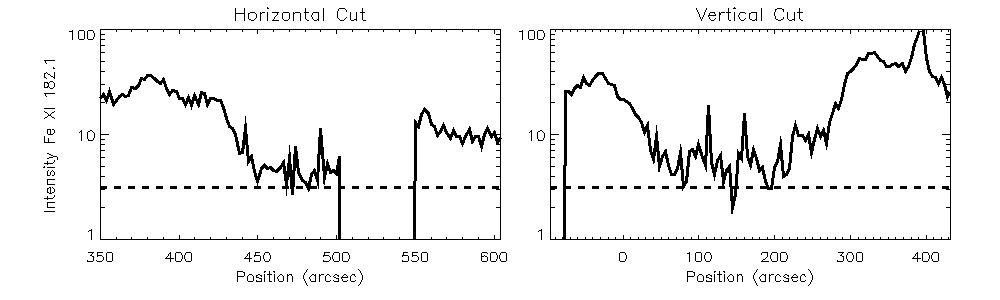}
\includegraphics[width=3.45in,angle=0]{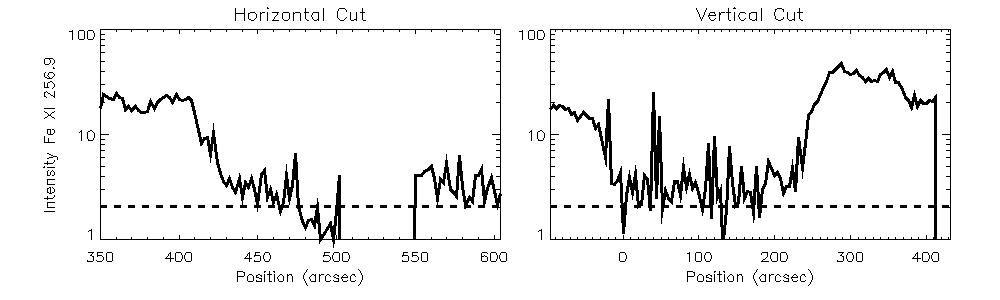}
\includegraphics[width=3.45in,angle=0]{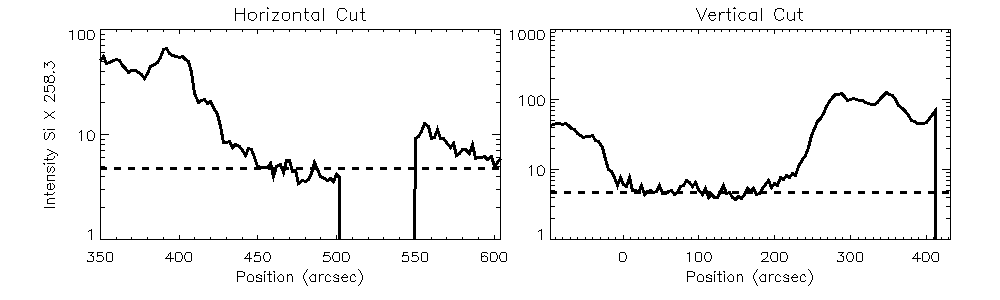}
\includegraphics[width=3.45in,angle=0]{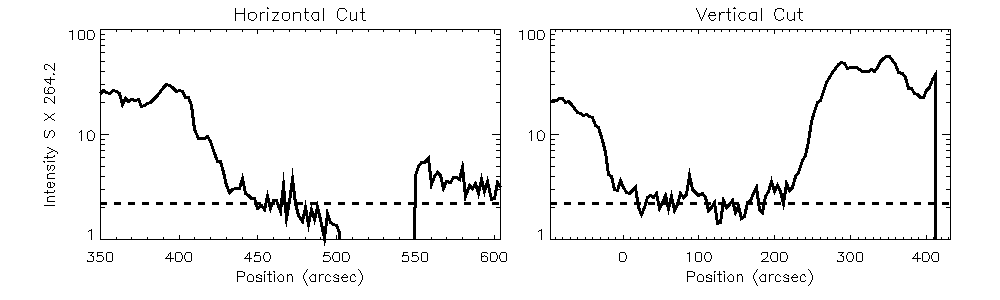}
\includegraphics[width=3.45in,angle=0]{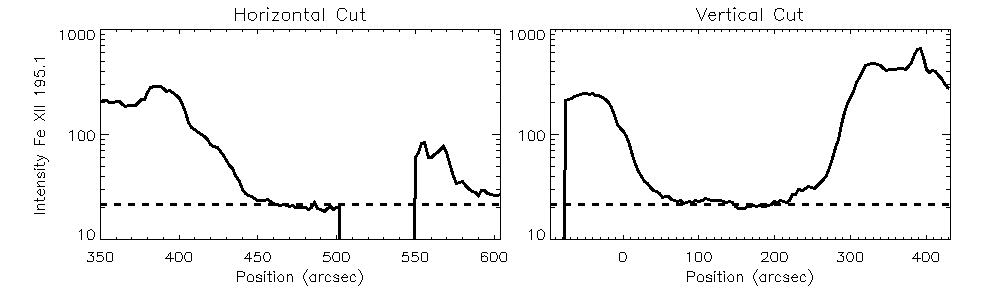}
\includegraphics[width=3.45in,angle=0]{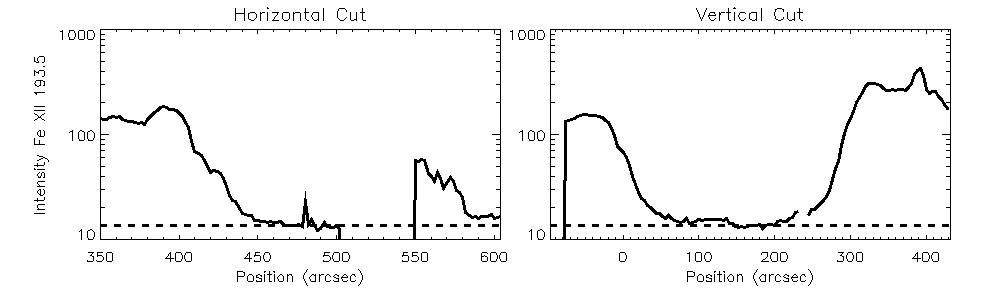}
\includegraphics[width=3.45in,angle=0]{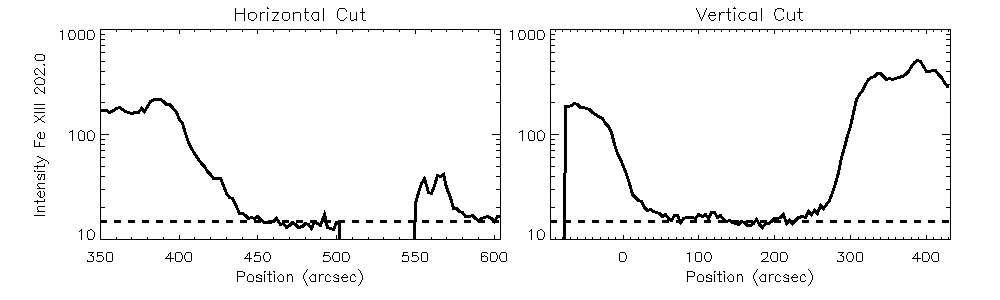}
\includegraphics[width=3.45in,angle=0]{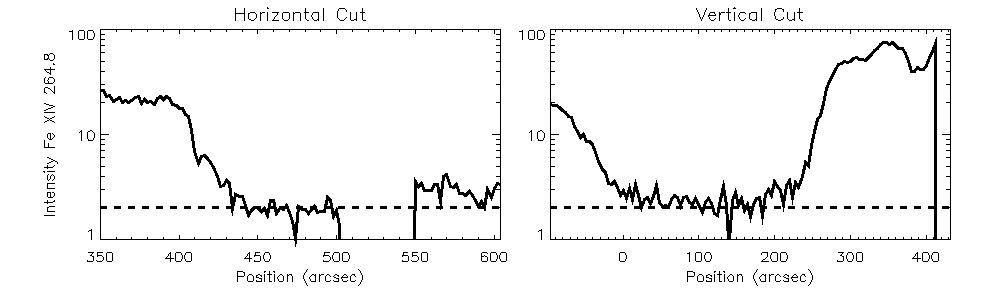}
\includegraphics[width=3.45in,angle=0]{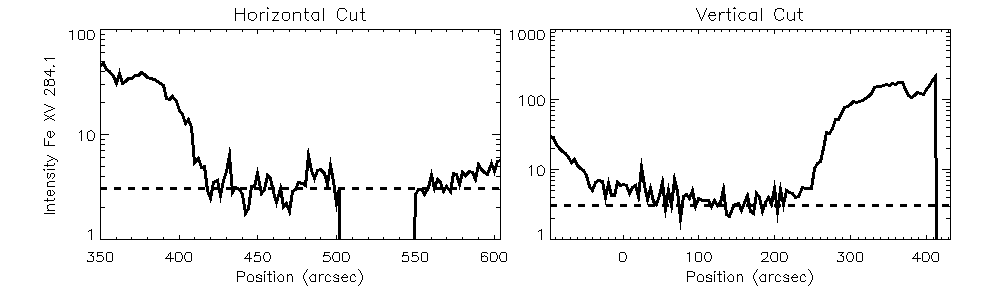}
\caption{Intensity cuts in both the horizontal and vertical direction for the spectral lines listed in Table 1 of our primary data set on 2007 March 31. The horizontal dashed line indicates 10\% of the median quiet Sun intensity for each spectral line.}
\end{center}
\end{figure*}

Beyond \ion[Fe x] the ratio of the coronal hole to quiet Sun intensity changes drastically behavior by 1) decreasing substantially to $\approx$10\%, and 2) becoming nearly constant for temperatures above Log~T=6.1. Both data sets show the same behavior. Furthermore, a close inspection of Figures 1 and 2 shows that the intensity map of lines formed above that temperature flattens out and no plasma structure is discernible above the noise level in all three measured quantities: line intensity, width, and centroid. These two results thus beg the question of whether the constant value of the intensity ratio at coronal temperature is due to the averaging of the intensities and masks a spatial variability, or to this ratio is really uniform across the whole coronal hole as Figures~1 and 2 seem to imply.

In order to answer this question we examined intensity "cuts" across selected X and Y positions in the intensity maps. We report in Figure 5 examples taken at positions 144" for SW and 124" for LW in the horizontal direction and at the position 470" for the vertical direction for our 16 spectral lines (in order of decreasing temperature). The dashed horizontal line represents 10\% of the median quiet Sun intensity for each line. Figure~5 shows that the intensity of the quiet Sun is at least one order of magnitude than the coronal hole intensity for the hottest lines in our dataset, and this vast difference in intensity diminishes as the ion temperature as the ion formation temperature decreases until it disappears for the "coolest" lines in out data set. Additionally, the coronal hole appears to lose all structure at higher temperature. The intensity of the coronal hole sits well above the 10\% line for cool lines (such as \ion[Fe viii] at 185.21 {\AA} and \ion[He ii] at 256.32 {\AA}); whereas the coronal hole intensity falls to a nearly constant value near the 10\% line for all ions formed at around Log T $\sim$ 6.15 K and above.

Figure 5 shows that the coronal hole intensity of the hot lines is flat, which is consistent with the lack of structure seen in all of the maps of Figure 1. In addition, the coronal hole intensity of the lines hotter than \ion[Fe xi] is around 10\% of the quiet Sun median values, regardless of their different formation temperatures. Ions with formation temperatures below 1 MK in the coronal hole intensity show variability and have different ratios to the quiet Sun from that of other ions.

In order to see if these results are consistent, we applied the same approach for our secondary data set. Figure 6 shows horizontal and vertical cuts across intensity at positions 501" in the horizontal direction and at the position 297" for the vertical direction. Again the lines are organized in order of decreasing temperature and the dashed horizontal line represents 10\% of the median quiet Sun intensity for each line. The lower signal to noise, and the lack of sufficiently high signal-to-noise for many of the lines in our list, forced us to show only the strongest lines. Despite the lower signal-to-noise, a similar trend as in the previous data set is observed. The coronal hole intensity dips down to 10\% of the median quiet Sun intensity for the hotter ions, but the coronal hole intensity is well above this line for cooler ions. Despite the fact that these are two different observation periods of separate coronal holes, both Figures 5 and 6 show the same behavior. The spatially flat and constant value of coronal hole intensity for hotter ions for both data sets suggests that this feature might be due to instrumental scattered light, and that this behavior did not change in the 6 years of the mission that separates the two data sets.

\vspace{2in}

\section{Expected intensities in coronal holes}

In the previous section we showed that there is a consistent 10\% value seen in coronal hole emission for our hotter ions. The intensity, centroid, and width maps lose any signature of structure inside coronal holes; the coronal hole to quiet Sun intensity ratio flattens to 10-15\% beyond a certain temperature, and the cuts show that all of these different lines are well above the 10\% level for cooler ions (despite their differences in formation temperature) and yet all of the hot lines look like nothing but noise for the coronal hole. However it is not clear yet what exactly is causing this pattern. In this section we discuss the potential sources of this 10\% value in hopes to determine the cause of this trend.

\subsection{Contamination from Closed Field Structures Along the Line of Sight}

The presence of measurable intensity from highly ionized species in an
equatorial coronal hole might be due to the presence of streamers along
the line of sight. However, though such a presence is likely at some
height along the line of sight, it is unlikely that these additional
structures provide such a high level of contamination. The reason is
twofold, both tied to the strong dependence of the electron density
on height that has been routinely observed in streamers.

First, streamer presence might induce some spatial modulation of the coronal
hole intensities that they would induce. This is because the equatorial
coronal hole we are considering has a non-negligible extension, so that
at any position inside the coronal hole, the overlying streamer will intersect
the line of sight at different heights from those of other areas of the
coronal hole. Since plasma emission is roughly proportional to the square
of the electron density, this would cause different amounts of streamer intensity to be added to the coronal hole emission, leading to a spatial variability
that we do not observe.

The second, and most important, reason is due to the fact that in order
to provide $\approx$10\% of the surrounding quiet Sun, the density of the
overlying streamer plasma must be relatively high (roughly a factor 3 less
than the one measured in the quiet Sun, assuming the same thermal structure as the quiet Sun).

Given the typical scale heights of the electron density, this means that the streamer
should intersect the line of sight within 0.1 solar radii from the surface, at any place
in the coronal hole. As a consequence, the magnetic field that confines the streamer
plasma should be very strongly bent towards the coronal hole, much more than observed
in eclipse images of large scale streamers.


\subsection{Coronal Hole Intensity Calculation}

In case real local emission is responsible for the observed intensities of
the hottest ions in the data set, namely \ion[Fe xiii-xv], its mere presence
indicates that significantly hot plasma is present along the line of sight,
that needs to be explained by theoretical models.

In order to test this idea, we have calculated the intensity from several strong lines from
Fe ions that would be expected from an equatorial coronal hole using
theoretical models that predict the plasma thermodynamic properties.
A number of models have been developed in the past, and recently
\citet{2014ApJ...790..111L} benchmarked three of them using a diagnostic
technique that compared their predictions with both in-situ measurements
of solar wind charge state distributions from Ulysses/SWICS and high resolution spectral line intensities observed by SoHO/SUMER at the solar limb.

\begin{figure*}[!ht]
\begin{center}
\includegraphics[width=3.45in,angle=0]{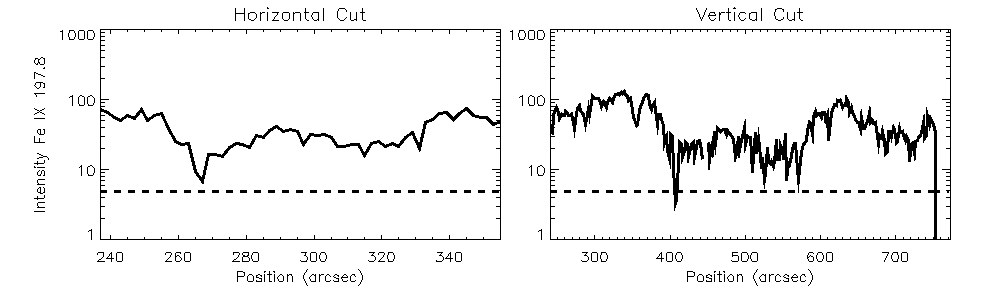}
\includegraphics[width=3.45in,angle=0]{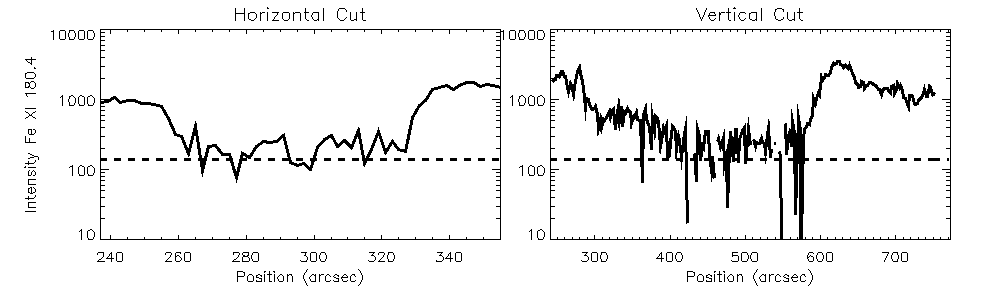}
\includegraphics[width=3.45in,angle=0]{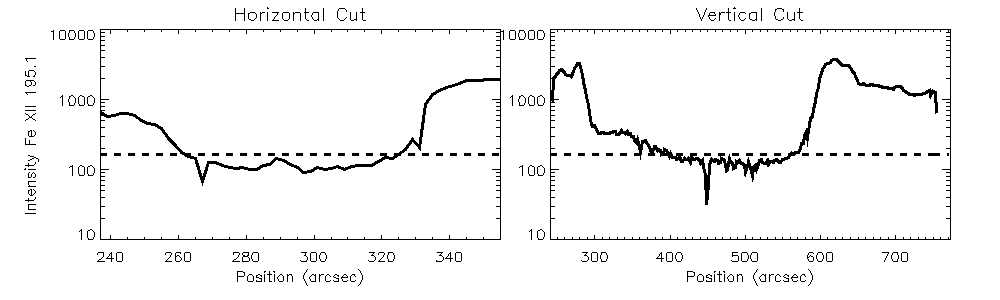}
\includegraphics[width=3.45in,angle=0]{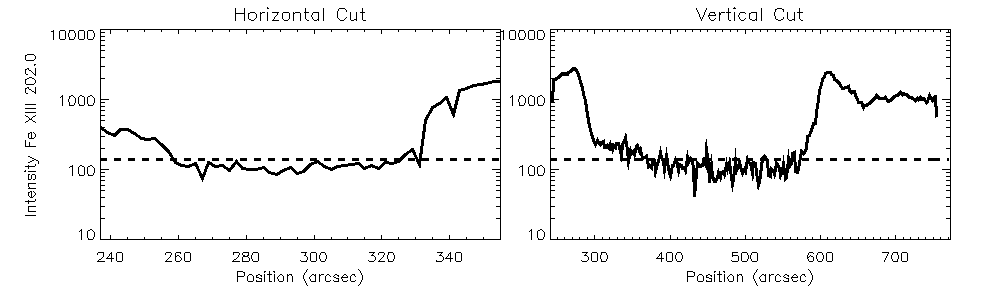}
\caption{Intensity cuts in both the horizontal and vertical direction for spectral lines of our second data set on 2013 October 12. The horizontal dashed line indicates 10\% of the median quiet Sun intensity for each spectral line.}
\end{center}
\end{figure*}

\citet{2014ApJ...790..111L} found that the models of \citet{2007ApJS..171..520C} --
1-dimension, 1-temperature model powered by waves -- and \citet{2014ApJ...782...81V} AWSoM model -- 3-dimension, 2-temperature model powered by
Alfven waves -- provided similar levels of agreement with observations.
Also, a comparison of these two models with electron density and
temperature measurements obtained with SUMER line intensity ratios
very close to the limb showed excellent agreement.

The latter result is very important, as the innermost (and densest) regions
of the corona are the most important for predicting line intensities
observed on the disk. Thus, we have carried out calculations of line
intensities for several key lines using the \citet{2007ApJS..171..520C} model
(hereafter C07); we have used their {\em equatorial} coronal hole
model since we are studying a data set observed near disk center.
Results from the 3D AWSoM model are qualitatively similar to those from
the C07 model, an unsurprising result given the similarity of their
model predictions close to the Sun, as noted by \citet{2014ApJ...790..111L}. Thus,
for the remainder of this work, we focus on the C07 model only.

The calculations have been done in several steps, in order to account for
the line of sight integration and for the departures of ion abundances
from equilibrium values induced by the wind acceleration, as described
by \citet{2012ApJ...744..100L}:

\begin{enumerate}

\item First, the Michigan Ionization Code (MIC -- \citet{2012ApJ...744..100L}) was
used to calculate the charge state evolution of the solar wind using
the C07 electron temperature, density and wind bulk speed predictions;

\item Second, we have calculated the emissivity of each volume element
along the solar wind trajectory using the non-equilibrium ion abundances resulting
from MIC, and the local electron density and temperature predicted by C07;

\item Third, we made an assumption about the line of sight and on the
plasma distribution along it and used it to calculate the final synthetic intensities.
\end{enumerate}

\noindent
Step~1 was carried out solving the system of equations

\begin{eqnarray}
\frac{\partial y_m}{\partial t} & = & n_e{\Big[{y_{m-1}C_{m-1}(T_e)+ y_{m+1}R_{m+1}(T_e)}\Big]} \nonumber \\
& &  + y_{m-1}P_{m-1} - \nonumber   \\
& & - y_m{\Big[{n_e{\Big({C_m(T_e)+R_{m}(T_e)}\Big)}+P_m}\Big]} \label{eq2} \\
\Sigma_m y_m & = & 1 \nonumber
\end{eqnarray}

\noindent
where the photoionization term, $P_m$, is given by

\begin{equation}
P_m = \int_{\nu_m}^{\infty}{\frac{4\pi J{\left({\nu}\right)}\sigma_m{\left({\nu}\right)}}{h\nu}d\nu}
\label{photoion}
\end{equation}

\noindent where $T_e$ is the electron temperature, $n_e$ is the electron
density, $R_m$, $C_m$ are the total recombination and ionization rate
coefficients from CHIANTI V8 \citep{1997A&AS..125..149D,2015A&A...582A..56D},
$y_m$ is the fraction of a given element in charge state $m$ normalized
to unity, $h$ is the Planck constant, $c$ is the speed of light, $\sigma_m$
is the photoionization cross section for the ion $m$, $\nu_m$ is the
frequency corresponding to the ion's ionization energy, and
$J{\left({\nu}\right)}$ is the mean spectral radiance of the Sun, in
units of erg~cm$^{-2}$s$^{-1}$sr$^{-1}$Hz$^{-1}$. $J{\left({\nu}\right)}$
was taken from TIMED/SEE \citep{2005JGRA..110.1312W} for 2007 March 31, 
the day of the main data set we use, while the photoionization cross 
sections have been taken from the parametric fits of \citet{1996ApJ...465..487V}.

In Step~3, we have assumed that the wind trajectory (e.g., the open magnetic 
field line along which the wind plasma is flowing) stretches vertically for 
the first 0.2 solar radii (the only region in the corona giving significant 
contribution to the total intensity) and is aligned with the line of sight.

In order to be meaningful, the calculated intensities need to be compared
to the predictions of a similar theoretical model. This is for two reasons:
first, models rely on a quantity of assumptions and approximations that
define the geometry, the physical processes included and those neglected
and so on; and second, models also rely on atomic physics to calculate
line and continuum intensity, which come with their own sets of 
uncertainties and approximations. Measured line emission suffers 
from none of these limitations.

However, C07 did not provide a model for the closed corona, and any other 
model that can be used will not be completely consistent with C07's assumptions, 
physical processes, and geometry, so that the comparison will not be completely 
unbiased and actually could introduce some more systematic effects due to the
inconsistencies between models. Atomic physics uncertainties, on the contrary,
will affect such model predictions in the same was as for C07 so that the
two sets of theoretical intensities will be biased in the same way and thus
will be comparable. On the other hand, predicting line intensities using a 
standard, unrelated DEM curve from another region will also be affected by
atomic physics in the same way, it will not introduce and model bias, but
will provide another type of inconsistency since these intensities can be
considered as semi-empirical.

Even though there is not perfect recipe for such a comparison, and given 
that we are interested not in an accurate comparison of predicted to observed
coronal hole-to-quiet Sun ratios but rather in its trend with temperature,
we chose to compare the C07 line intensities with those obtained with the 
"EIS" DEM curve available in the CHIANTI package in SolarSoft. This DEM was 
determined by \citet{2009ApJ...705.1522B} over a completely unrelated quiet Sun 
region close to disk center, utilizing a fairly large amount of ions and 
lines, so that systematic errors given by a particular spectral line
were offset by the presence of many other lines whose atomic physics was
more accurate. While, again, the use of a DEM provides some level of
inconsistency when intensity ratios are calculated using coronal hole
intensities coming from a fully theoretical model, it has the great 
advantage of providing a relatively accurate prediction of the quiet Sun
intensity of coronal lines at temperatures higher than log~T$\simeq 6.15$, 
which is the range that most interest us; purely theoretical models of the 
quiet Sun, by relying on a still unknown coronal heating mechanism, may 
still provide levels of emission significantly different from the 
observed ones.

Figure 2 shows that the observed ratio between coronal hole and
quiet Sun intensities reaches a constant value at temperatures larger
than Log T $\approx 6.1$ K. In order to test whether our calculations
reproduce such a behavior, we compare the measured ratios in Figure 2 with 
the ratios between the predicted equatorial coronal hole intensities and 
the measured quiet Sun.

\subsection{Results}

\begin{figure}[b]
\includegraphics[width=8.5cm,angle=0]{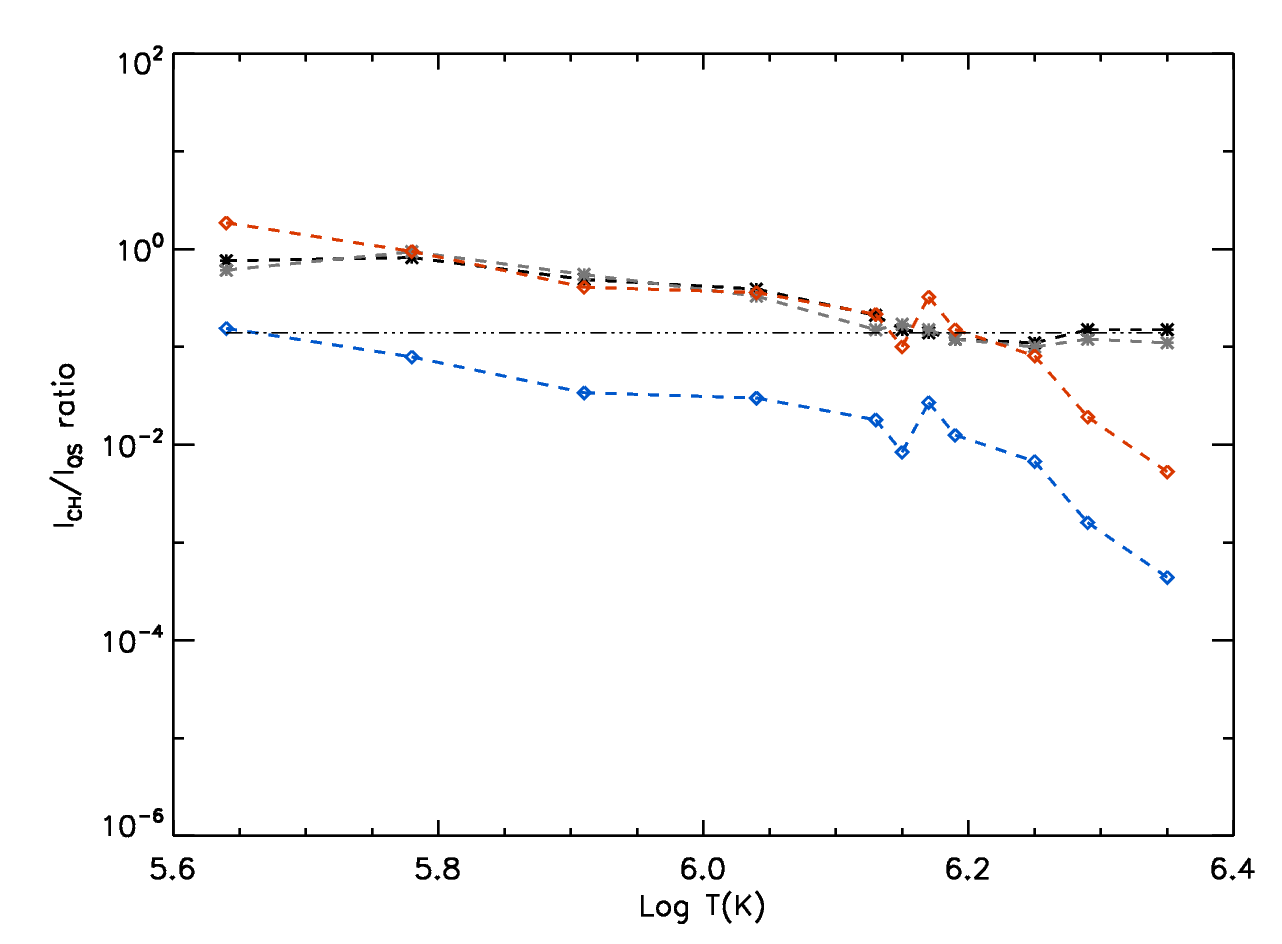}
\caption{Observed (black and grey dashed lines) and predicted (blue dashed line)
intensity ratio for the ions listed in Table 2.
The red curve shows the synthetic ratios, multiplied by a factor 12
to match the observed values below Log T=6.2 K. The horizontal line
indicates the 0.15 level.}
\label{ch_qs_ratio_2.pdf}
\end{figure}

The equatorial coronal hole to quiet Sun intensity ratios are listed in
Table 2. The \ion[He ii] 256.3 {\AA} line has been omitted from Figure 7 (as well as from Figure 2) because this line is optically thick and its intensity can not be calculated accurately with the approximations we have used in this work.
The absolute value of these ratios indicates that the predicted 
equatorial coronal hole line intensities are much lower than their observed 
counterparts, but comparable to them. 

Figure 7 shows the same measured coronal hole to quiet
Sun intensity ratios (black and grey dashed lines), compared to the calculated ones
(blue dashed line). A horizontal dash-dotted line at 0.15 is drawn to
approximately indicate the level of the ratio values at temperature larger
than Log T = 6.15 K. The theoretical ratios at temperatures below Log T = 6.15 K
are lower by a factor $\approx$10-15 from the observed ones; the red dashed
line in the figure simply increases the calculated ratios by an arbitrary
factor~12  to bring the calculated and observed curves together at sub-coronal
temperatures and visually compare their behavior. The underestimation of the
synthetic coronal hole intensities, is caused by 1) the approximations taken
in our calculations; 2) the differences between the specific data set we
have used and the characteristics of the coronal holes used by \citet{2007ApJS..171..520C} to develop their model; and 3) to possible inaccuracies in
the model itself. Nonetheless, it is remarkable that a solar wind model
is able to provide intensities so close to the values of a completely
unrelated observation, after all the approximations we made are considered.

The most important feature of Figure 7 is that it shows
that the temperature dependence of the predicted ratios is fairly similar
to the observed one until \ion[Fe xi] (formed at Log T = 6.13 K), whose position in the red curve
lies however at exactly the 0.15 line we drew. At temperatures higher than
those of this ion, the predicted ratios decrease far below their observed
counterparts.

The reason why the predicted ratios decrease so strongly
can be understood by considering the behavior
of the solar wind electron density and temperature evolution on one side,
and of its charge state dependence on the other. First, the solar wind
electron temperature rises very quickly from chromospheric to sub-coronal
values (just short of 1~MK) but then rises very slowly to reach 1.3~MK at
around 1.3 solar radii (see Figure 8). At this height, where non-negligible amounts of
\ion[Fe xiii-xv] could form under ionization equilibrium, the electron
density is already decreased to around $10^7$~cm$^{-3}$, causing their
emission to be very low.

\begin{table*}
\begin{center}
\begin{tabular}{lllrrccrrcc}

Ion & $\lambda$ ({\AA}) & EQ & \multicolumn{4}{c}{2007 March 31} & \multicolumn{4}{c}{2013 October 12} \\
& & & ECH & QS & ECH/QS Ratio & EQ/QS Ratio & ECH & QS & ECH/QS Ratio & EQ/QS Ratio \\ [0.5 ex] 
\\
\hline \\
\ion[He ii] & 256.32 & \nodata & 133.6 & 232.2 & 0.575 &\nodata &\nodata &\nodata &\nodata & \nodata \\
\\
\ion[Fe viii] & 185.21 & 9.22 & 16.4 & 21.7 & 0.756 & 0.425& 66.0 & 108.7 & 0.607 & 0.085 \\
\\
\ion[Si ii] & 275.37 & 2.99 & 10.1 & 12.3 & 0.821 & 0.243& 26.3 & 27.9 & 0.942 & 0.107 \\
\\
\ion[Fe ix] & 198.86 & 0.876 & 4.61 & 9.38 & 0.491 & 0.093& 28.1 & 48.0 & 0.585&0.018\\
\\
\ion[Fe x] & 177.23 & 11.8 & 51.5 & 140.1 & 0.368& 0.084& \nodata & \nodata & \nodata & \nodata\\
\\
\ion[Fe x] & 257.27 & 1.97 & 17.9 & 44.1 & 0.406& 0.045&\nodata &\nodata & \nodata &\nodata \\
\\
\ion[Fe xi] & 180.40 & 8.62 & 48.3 & 240.2 & 0.201& 0.035& 323.0 & 1404.0 & 0.230 &0.006\\
\\
\ion[Fe xi] & 182.17 & 0.566 & 6.61 & 31.1 & 0.213 &0.018 & 27.8 & 213.7 &0.130 &0.003\\
\\
\ion[Fe xi] & 256.93 & \nodata & 4.42 & 20.5 &0.216 & \nodata &\nodata &\nodata & \nodata &\nodata\\
\\
\ion[Si x] & 258.38 & 0.317 & 6.90 & 47.2 &0.146 &0.007 & 32.3 & 189.2 & 0.171 & 0.002\\
\\
\ion[S x] & 264.24 & 0.202 & 3.13 & 22.2 & 0.141&0.009 & 11.1 & 72.8 &0.152 &0.003 \\
\\
\ion[Fe xii] & 195.12 & 1.71 & 26.3 & 214.0 & 0.123 &0.008 & 202.9 & 1652.0 & 0.123 & 0.001\\
\\
\ion[Fe xii] & 193.51 & 1.06 & 16.5 & 135.2 &0.122 &0.008 & \nodata & \nodata & \nodata & \nodata\\
\\
\ion[Fe xiii] & 202.05 & 0.299& 16.8 & 149.0 &0.113 & 0.002& 136.6 & 1383.0& 0.099& $2.16 \times 10^{-4}$\\
\\
\ion[Fe xiv] & 264.79 &$7.73 \times 10^{-3}$ & 2.92 & 20.0 & 0.146& $3.87 \times 10^{-4}$&\nodata &\nodata &\nodata &\nodata \\
\\
\ion[Fe xv] & 284.17 & $4.61 \times 10^{-3}$ & 4.61 & 30.7 &0.150 & $1.50 \times 10^{-4}$ &\nodata &\nodata &\nodata &\nodata \\
\\
\hline \\
\end{tabular}
\end{center}
\caption{Predicted (EQ) and measured (ECH, QS) intensities for select spectral lines. ECH stands for Equatorial Coronal Hole, QS for quiet Sun. EQ predicted intensities have been calculated using the equatorial coronal hole model of \citet{2007ApJS..171..520C}.}
\label{synthetic_ratios}
\end{table*}

On the other hand, the amount of \ion[Fe xiii-xv] ions formed at
this height is lower than expected from the equilibrium values, because
the solar wind is under-ionized due to the so-called ``delay effect''
\citep{2012ApJ...761...48L}. This effect, which causes the wind ionization to be
lower than equilibrium values at the local temperature, is due to the
fact that as the solar wind accelerates from the chromosphere to the
corona, the plasma charge state distribution does not have enough time
to adjust to the higher local temperature at any location because its speed causes it
to dwell at each location less than its equilibration time. This causes
the overall wind charge state distribution to be left to values typical
of lower temperatures. This effect, eventually leading to
the wind plasma freeze-in, is already present in the transition region
and propagates in the corona, further depressing the emission of the
most highly ionized species of any element. For example, \ion[Fe xiii-xv]
ion abundances are depressed by a factor 1.4 to 2 from their ionization
equilibrium values below 1.1 solar radii, where their emission is
predicted to be largest.

These arguments provide support to the idea that the intensities of
the most highly ionized species present in the coronal hole data set
we have analyzed are not due to local emission, but rather to the
presence of significant scattered light.

\vspace{2in}

\section{Discussion}

\subsection{Scattered light fraction}

In the previous sections we have argued that the observations made in
an equatorial coronal hole are affected by a significant amount of
scattered light, which we quantified to be around 10-15\% of the
intensity of the surrounding quiet Sun. This amount is larger than
indicated by UU10, and this comes
as no surprise. In fact, UU10 estimated the amount of scattered light
using observations carried out while the EIS spectrometer was observing
a partial eclipse of the solar disk. The whole argument of that study was that any
intensity observed by EIS in the occulted part of the corona was only
due, after dark current subtraction, to instrument-scattered light
emitted by regions not occulted by the Moon.

\begin{figure}[t]
\includegraphics[width=4.9cm,angle=90]{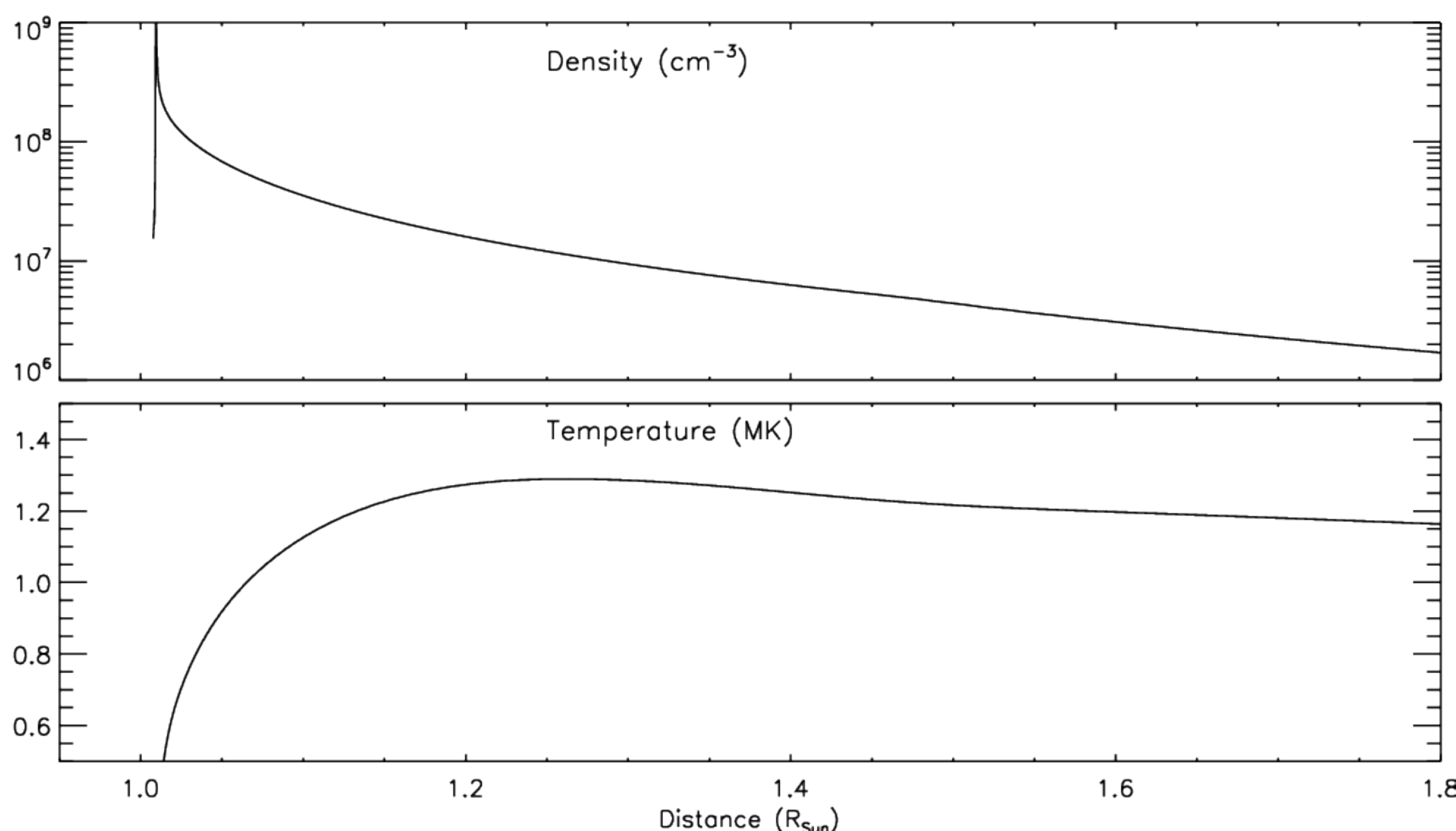}
\caption{Electron density (\textbf{top}) and temperature (\textbf{bottom}) versus height above the limb (at 1.0 R$_{\odot}$) predicted by the \citet{2007ApJS..171..520C} equatorial coronal hole model.}
\label{}
\end{figure}

The observational setup of the UU10 observation is qualitatively
similar to observations made at the solar limb, where the EIS telescope
is exposed to significant radiation only for the portion of its field
of view that includes the solar disk and the closest regions of
the solar limb. In the case of observations of coronal holes made
inside the solar disk the situation is different, especially when
far from the limb, because in these cases the entire field of view
of the telescope is illuminated. This accounts for the higher fraction
of quiet Sun intensity that is scattered into the coronal hole itself, 
as measured from our work.

It should be noted that our measurement of instrumental scattered light 
equally affects both the coronal hole and quiet Sun region. Although this 
scattered light moreso affects the coronal hole due to its low intrinsic 
signal, the scattered light signal at any point in the FOV is a convolution 
of all of the emission over the disk. Moreover, the level of scattering 
depends on the surrounding features seen on-disk. Therefore, the amount 
of scattered light that we have measured can in principle be smaller 
than the contamination occurring where active regions are present, and 
even less when EIS observed flares. 

As previously noted in other instruments, the primary scattering mechanism 
is due to scattering off of the mesh supporting the entrance filters 
(e.g. in SDO/AIA -- \citep[e.g.,][]{2001SoPh..198..385L,aiapsf,2013ApJ...765..144P}). 
In these cases the PSF is dominated by the filter diffraction pattern off 
of the mesh. It is likely that the scattering which we observe with EIS is 
due to this phenomena, which urges for a similar PSF for EIS. Indeed, such 
mesh causes the scattered intensity to form a diffraction pattern which is 
particularly evident in flare images taken by those instruments. We visually 
inspected a few EIS observations of several large flares, and we indeed 
found a qualitatively similar pattern. A couple of examples are shown 
in Figure~9, which shows the diffraction pattern observed with the 
\ion[Fe xxiv] 255.1~\AA\ line observed during the 10 September 2017 
flare at 16:18~UT (left panel), and even in the \ion[Fe xv] 284.1 line 
for our secondary data set (right panel). It is interesting to see that 
the flare image even shows at (1080",-270") the second peak at $x=1.5\pi$ 
of the diffraction spike series as described in \citet{aiapsf}.

\begin{figure*}
\begin{center}
\includegraphics[width=8.0cm,height=10.0cm]{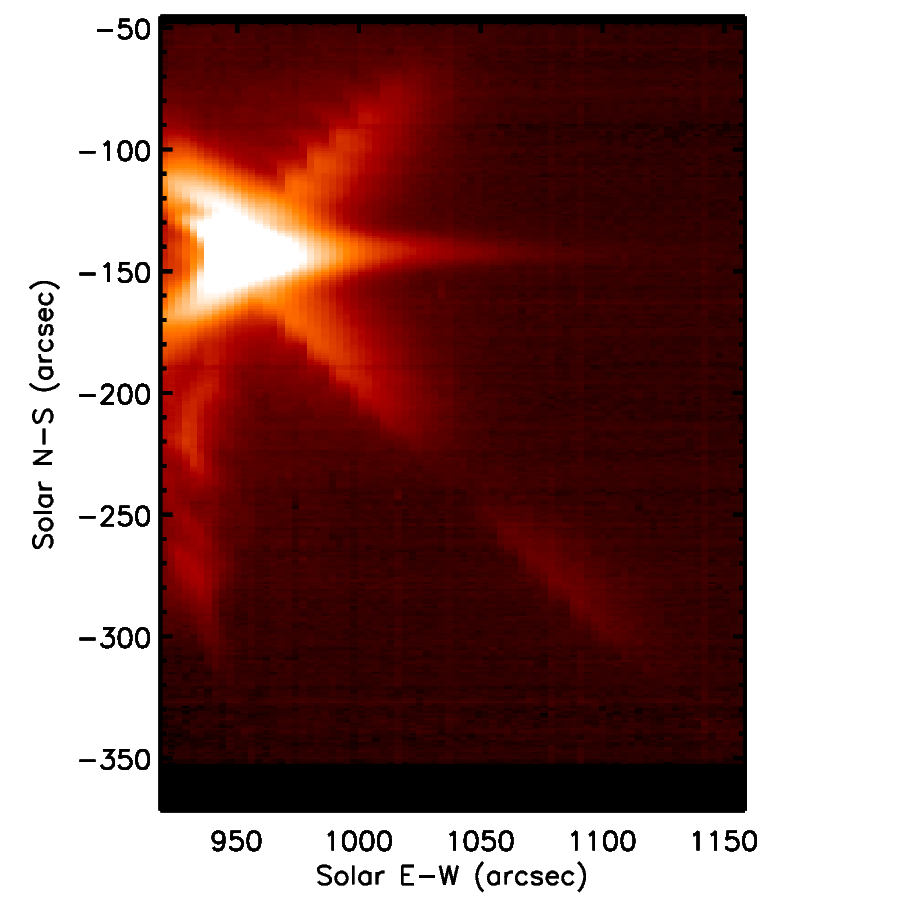}
\includegraphics[width=8.0cm,height=10.0cm]{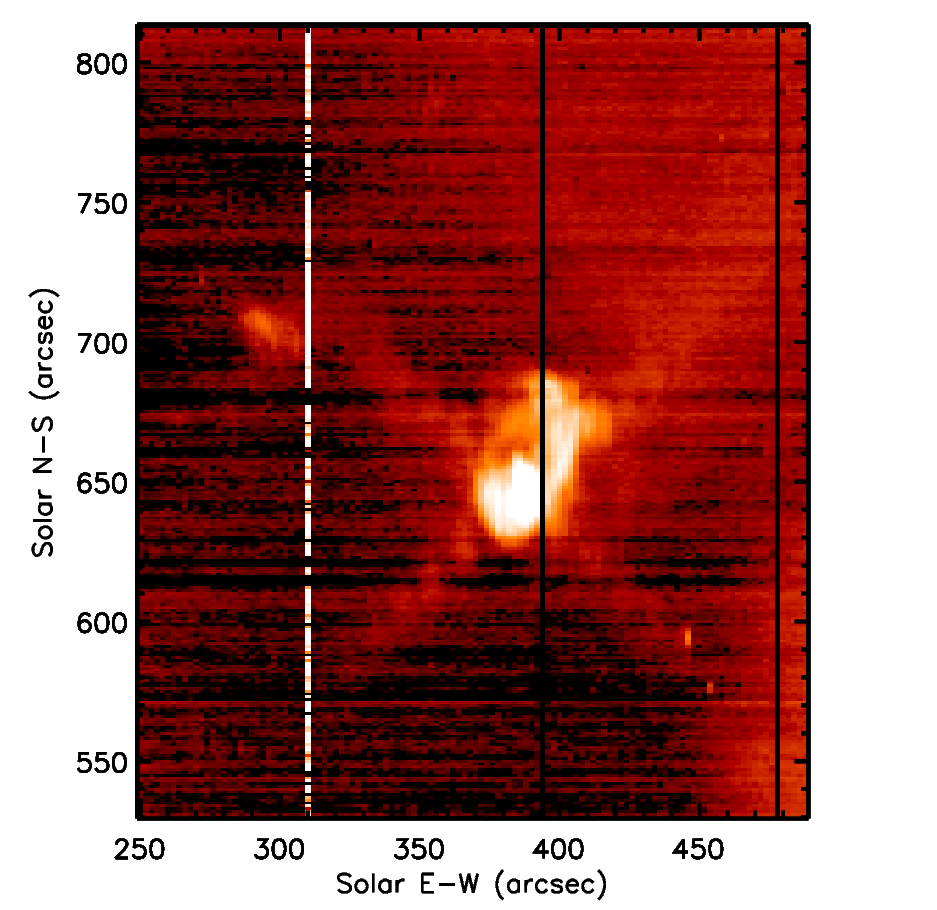}
\caption{ Intensity maps of the \ion[Fe xxiv] 255.1~\AA\ line observed in the
10 September 2017 X8.2 flare at 16:18~UT {\bf (left)}, and of the \ion[Fe xv] 284.1~\AA\
line observed in our secondary data set {\bf(right)}.}
\end{center}
\end{figure*}

These arguments beg for a more systematic study of instrument-scattered
light both as a function of position of the coronal hole in the disk,
also taking into account the presence of active regions and their
distance from the coronal hole. In fact, active regions serve as
localized regions of strongly enhanced intensity and could easily
introduce an additional contribution of scattered light which depends
on the distance from the active region itself.

\vspace{2in}

\subsection{Coronal hole plasma diagnostics}

The presence of a substantial amount of scattered light in coronal
hole observations poses a significant hurdle in the measurement of
the coronal hole's physical properties. In fact, the 10-15\% of
the quiet Sun intensity which constitutes scattered light affecting
un equatorial coronal hole turns out to be a large, if not dominant
component of the measured coronal hole line intensity for the majority 
of coronal ions. In fact, while
in the upper transition region (\ion[Fe viii], \ion[Si vii]) such a
contribution is relatively minor and unlikely to fundamentally change
plasma diagnostic results, for coronal ions it can easily account for
the majority of the observed photons, or even provide the whole of
the observed intensity altogether, like in the case of \ion[Fe xiv-xv].

The presence of this scattered light, if not accounted properly, can
easily lead to errors, such as, to name a few:

\begin{enumerate}

\item It may lead us to conclude that an excessive amount of hot plasma is present
along the line of sight (providing ions such as \ion[Fe xiv-xv]);

\item May alter measurements of the distribution of the plasma with
temperature (i.e. the DEM), and thus
significantly affect the calculation of synthetic spectra and of
radiative losses;

\item Substantially alter the results of plasma diagnostics through
line intensity ratios, one of the most used methods to infer the
plasma electron density and temperature in coronal hole plasmas.

\end{enumerate}

\noindent

Likewise, the presence of substantial contamination in the intensity of
coronal lines is likely to alter significantly the values of the centroids
and widths of the coronal hole lines. This is particularly important,
because it can lead to incorrect measurements of blue-shifts that can
be used to study the accelerating slow wind, while line widths can help us draw conclusions on ion heating in the wind plasma: while it is difficult
to assess a priori what the final effect would be on each individual
coronal line and data set, scattered light contamination casts shadows on any conclusions drawn from centroid measurements.

\section{Conclusions}

In this work we studied the intensities of hot lines formed in two equatorial coronal holes observed in the solar disk in two different moments along the solar cycle. Our goal was to quantitatively assess the importance of instrumental scattered light. We find that in equatorial coronal holes, for lines with a formation temperature larger than Log T $\sim$ 6.1 K:

\begin{enumerate}

\item Line widths and centroids have constant values throughout the coronal hole;

\item Line intensities also have constant values, showing no structure;

\item The measured coronal hole / quiet Sun ratios are constant at around 10-15\% regardless of position in the coronal hole;

\item This ratio is approximately the same in the 2007 and the 2013 data sets, indicating no change during the mission.

\end{enumerate}

We calculate the predicted coronal hole intensities and find that for lines with formation temperatures Log T $\sim$ 6.1 K or larger the predicted intensity is far lower that observed. We rule out the presence of streamer contamination as the explanation for the observed intensity values for these coronal lines. We discuss the importance of this contamination for plasma diagnostics in coronal holes, as well as why it is larger than the estimate previously provided by the EIS team. We note however that, given the observational configuration of the latter, the current estimate is still valid for off-disk plasmas in the vicinity of the solar disk.

An important consequence of this study is that coronal hole plasma temperature are likely much lower than usually estimated, in the sense that the high-temperature tails of the DEM determined in several studies might actually be an artifact of scattered light contamination. This has important consequences for theoretical models of coronal hole structure and solar wind acceleration.

\acknowledgements

The authors warmly thank Drs. Harry Warren and Ignacio Ugarte Urra (Naval Research Lab)
for suggestions and very helpful discussion concerning EIS scattered light, as well as
the anonymous referee whose insightful comments helped us to improve and clarify the
manuscript considerable. This work was supported by NASA grants NNX17AD37G and NNX15AB73G 
and NSF grant AGS-1408789. Hinode is a Japanese mission developed and launched by ISAS/JAXA, 
with NAOJ as domestic partner and NASA and UKSA as international partners. It is operated 
by these agencies in co-operation with ESA and NSC (Norway). CHIANTI is a collaborative 
project involving George Mason University, the University of Michigan (USA) and the 
University of Cambridge (UK).

\bibliography{Stray_light}

\end{document}